\documentclass[sigconf,balance=false]{acmart}
\usepackage{popets}
\usepackage{multirow}
\usepackage{todonotes}
\usepackage{xspace}
\usepackage{xcolor}
\usepackage{mdsymbol}

\acmConference{}

\newcommand{\tk}[1]{\todo[inline,color=red]{TK - #1}}

\newcommand{\dd}[1]{\todo[inline,color=green]{DD - #1}}

\newcommand{\numexperiments}{{1171}\xspace}
\newcommand{\numqueries}{{24530}\xspace}
\newcommand{\nummonths}{{20}\xspace}

\begin{document}

\title[]{Echoes of Privacy: \\ Uncovering the Profiling Practices of Voice Assistants}


\author{Tina Khezresmaeilzadeh}
\affiliation{%
  \institution{University of Southern California}
  \city{Los Angeles}
  \state{California}
  \country{USA}}
\email{khezresm@usc.edu}

\author{Elaine Zhu}
\affiliation{%
  \institution{Northeastern University}
  \city{Boston}
  \state{Massachusetts}
  \country{USA}}
\email{zhu.lae@northeastern.edu}

\author{Kiersten Grieco}
\affiliation{%
  \institution{Northeastern University}
  \city{Boston}
  \state{Massachusetts}
  \country{USA}}
\email{grieco.k@northeastern.edu}

\author{Daniel J. Dubois}
\affiliation{%
  \institution{Northeastern University}
  \city{Boston}
  \state{Massachusetts}
  \country{USA}}
\email{d.dubois@northeastern.edu}

\author{Konstantinos Psounis}
\affiliation{%
  \institution{University of Southern California}
  \city{Los Angeles}
  \state{California}
  \country{USA}}
\email{kpsounis@usc.edu}

\author{David Choffnes}
\affiliation{%
  \institution{Northeastern University}
  \city{Boston}
  \state{Massachusetts}
  \country{USA}}
\email{choffnes@ccs.neu.edu}


\renewcommand{\shortauthors}{Khezresmaeilzadeh et al.}

\begin{abstract}

Many companies, including Google, Amazon, and Apple, offer voice assistants as a convenient solution for answering general voice queries and accessing their services. 
These voice assistants have gained popularity and can be easily accessed through various smart devices such as smartphones, smart speakers, smartwatches, and an increasing array of other devices. However, this convenience comes with potential privacy risks. 
For instance, while companies vaguely mention in their privacy policies that they may use voice interactions for user profiling, it remains unclear to what extent this profiling occurs and whether voice interactions pose greater privacy risks compared to other interaction modalities.

In this paper, we conduct \numexperiments{} experiments involving a total of \numqueries{} queries with different personas and interaction modalities over the course of \nummonths{} months to characterize how the three most popular voice assistants profile their users. 
We analyze factors such as the labels assigned to users, their accuracy, the time taken to assign these labels, differences between voice and web interactions, and the effectiveness of profiling remediation tools offered by each voice assistant.
Our findings reveal that profiling can happen without interaction, can be incorrect and inconsistent at times, may take several days to weeks for changes to occur, and can be influenced by the interaction modality.

\end{abstract}


\maketitle

\section{Introduction}

Voice assistants, including Amazon Alexa~\cite{alexa}, Google Assistant~\cite{googleassistant}, and Apple Siri~\cite{siri}, have rapidly gained popularity as versatile tools for accessing information and services, reducing the need for traditional physical interfaces, with more than 142 million users in the US alone~\cite{statista-va-users}.
Despite their convenience, voice assistants still raise concerns regarding user privacy, mirroring issues found in other non-vocal interaction modalities such as web browsing. 
Previous research~\cite{barford-www-2014,bashir-2016-usenix,farooqi2020canarytrap,cook2020inferring,bashir2016tracing,carrascosa15conext,lecuyer2015sunlight,agarwal2020www} has demonstrated that online services accessed through web browsers or mobile applications often leverage user data for profiling, aiming to tailor personalized and sponsored content. 
Additionally, recent findings have revealed that certain interactions with third-party skills of the Alexa voice assistant are utilized for profiling purposes~\cite{iqbal-imc23}.
However, as far as we know, it remains uncertain whether and how voice assistants themselves, without utilizing third-party skills, engage in user profiling. Additionally, the extent to which their profiling methods resemble or differ from other interaction modalities, along with the associated privacy implications, remains unclear.

This highlights the necessity for further privacy research to explore this topic and offer transparency regarding how voice assistants profile their users. 
Profiling information becomes integrated into the user's data and can have various ramifications, including impacting the user experience by potentially leading to the display of irrelevant advertisements. 
Additionally, it can influence sponsoring contracts, such as determining inclusion or exclusion from relevant promotional campaigns. Moreover, there's a risk of unwanted or incorrect information being shared with third parties, highlighting the importance of understanding and addressing the potential side effects of user profiling in this new context.

The goal of this research is to shed light on the privacy implications due to interactions with voice assistants. 
This includes investigating the scope, accuracy, and timing of profiling, such as what aspects are being profiled and whether this profiling is pertinent. 
Additionally, the research will explore differences in profiling between voice interactions and other modalities, such as web browsing, to understand if and how the methods diverge.
Finally, we will analyze the effectiveness of the profiling mitigation tools offered by each voice assistants such as opt-out mechanisms and the possibility to delete or rectify incorrect labels.

Unfortunately, attaining such a goal is challenging because the most popular voice assistants are not software that runs on users' premises, but instead they operate as cloud services, making it very difficult to audit their internal operations. 
While the privacy policies of these platforms disclose the potential for user profiling based on voice interactions~\cite{google-privacy-policy,amazon-privacy-policy,apple-advertising}, they typically lack  details on the methodology involved. Fortunately, privacy regulations in certain jurisdictions (e.g., GDPR~\cite{gdpr} and CCPA~\cite{ccpa}) compel these platforms to allow users to make data disclosure requests~\cite{google-myadcenter,amazon-adprefs,apple-privacy}. 
Through these requests, platforms are required to provide users with all the information they have collected or generated, including user profiles, often described in terms of demographic (e.g., income group) and interest labels (e.g., an interest in cars).

To address the challenge above and achieve our goal, we developed a rigorous experimental methodology and applied it to the three most prevalent voice assistant platforms: Google Assistant~\cite{googleassistant}, Alexa~\cite{alexa}, and Siri~\cite{siri}. Our approach involves the creation of new user accounts, each meticulously trained with a curated set of voice queries designed to simulate various user personas, each one with an expected set of target profiling labels.
Then, we leverage the data disclosure features inherent to each voice assistant platform to monitor the appearance of actual profiling labels. 
For instance, if an account is trained to mimic a ``luxury persona'' with questions tailored around luxury topics, we anticipate the emergence of labels aligned with this persona, such as ``high income.''

After applying this methodology to the three platforms above and completing \numexperiments{} experiments with a total of \numqueries{} voice queries over the course of \nummonths{} months, we discovered that Apple Siri does not produce profiling labels, while in the remaining platforms profiling does occur and can be erroneous and inconsistent at times.
For example, we have seen situations in which profiling happens without any user interactions, resulting in arbitrary profiling labels (Google assistant).
Moreover, the process of profiling varied significantly in accuracy and duration, with unexpected labels appearing (Google assistant), different labels following the same questions at different times (Google assistant), and profiling labels taking days to appear (Google assistant and Amazon Alexa).
Importantly, we observed that the interaction modality can influence the accuracy and timing of profiling, with different profiling outcomes depending on whether the interaction occurs through via voice commands or via web (Google assistant and Amazon Alexa).
Finally, we tested the mitigation tools each platform offered and whether any label rectification or opt-out request was correctly fulfilled.

\noindent Summarizing, the contributions of this paper are the following:
\begin{enumerate}
\item a new persona-based methodology for producing profiling labels in different voice assistant platforms (Section~\ref{sec:Methodology}),
\item a determination of whether a platform pre-populates its profiling labels on fresh accounts (Section~\ref{sec:Exploring_Prepopulated_Labels_in_Profiles}),
\item a characterization of voice assistant first-party profiling across several dimensions, such as accuracy and timing (Section \ref{sec:Experimental_measurement_of_profiling_activities}),
\item a comparison of profiling across different interaction modalities (Section \ref{sec:impact_of_interaction_modality_on_user_profiling}), 
\item an assessment of the efficacy of mitigation mechanism such as profiling opt-out and label rectification (\ref{sec:Mitigating_Privacy_Risks_in_Voice_Assistant_Profiling}), and
\item a discussion of the privacy implications of our findings (Section~\ref{sec:discussion}).
\end{enumerate}

Should this paper be accepted, we are committed to publishing all the code and dataset on GitHub. This will enable further research and facilitate comparisons with our findings.

\section{Related Work}
\label{sec:relatedwork}
\noindent \emph{Voice assistant privacy.}
Voice assistants and the devices typically used to access them (e.g., smart speakers), since their debut, have been affected by privacy and security concerns, as discussed in several recent surveys~\cite{li2023survey,cheng22survey,yan2022survey}. 
A large amount of existing privacy research focuses on risks that are likely due to technological limitations such as smart speakers misactivations~\cite{alexa_misactivation_real_world,dubois2020speakersPETS}, unintentional software bugs that sometimes led to unauthorized sharing of voice recordings~\cite{alexa_data_shared_with_stranger,smart_speaker_eavesdropping,googlehomeminiflaw17}, and other instances of data exposure due to third-party apps or skills~\cite{Cheng2020SkillPolciyViolationCCS,Young22SkillDetectiveUSENIX,iqbal-imc23}.
Another line of work has shown that it is possible to infer user labels from the characteristics of the recording instead of just from its transcription~\cite{Singh2019VoiceProfiling}, thus creating more profiling opportunities for voice assistant vendors.
While these works focus on important risks, we are complementary with respect to them since we focus on the measurement and mitigation of intentional data collection in terms of profiling labels to the first-party (i.e., the voice assistant vendor). Specifically, we focus on targeting the labels of each specific platform using explicitly biased utterances while maintaining voice characteristics as neutral as possible.

\vspace{-0.1in}
\paragraph{Persona-based inference of labels.}
In this work, we use a persona-based methodology to produce labeling in the context of voice assistants and then use data disclosure requests to obtain user labels.
This methodology was inspired by the ones used in a large body of prior work (e.g.,~\cite{barford-www-2014,bashir-2016-usenix,farooqi2020canarytrap,cook2020inferring,bashir2016tracing,carrascosa15conext,lecuyer2015sunlight,agarwal2020www}), where different synthetic profiles (personas) are used to study how data is propagated to third parties in contexts such as the web and mobile apps.
These works typically use personas to determine what data is shared with third parties~\cite{farooqi2020canarytrap}, server-side data exchanges~\cite{cook2020inferring,bashir2016tracing}, and other online tracking and advertising behavior~\cite{carrascosa15conext,lecuyer2015sunlight,agarwal2020www}.
Although our work also uses a persona-based approach, a limitation of most existing work is that it focuses on limited modalities (e.g., the web) that do not include voice assistant. One recent persona-based work~\cite{iqbal-imc23} focuses on voice assistants, but it has a different scope with respect to ours since it is limited to third-party skills, it only focuses on the Alexa voice assistant, it does not target the voice assistant vendor itself (i.e., the first party), and obtains labeling data indirectly and in a probabilistic way through the analysis of the ads ecosystem rather than using deterministic data disclosure requests as we do.

\vspace{-0.1in}
\paragraph{Other labeling determination approaches.}
Prior work in the broader context of Internet of Things (IoT), including smart speakers, has traditionally addressed the labeling problem by understanding what data is shared to their cloud platforms (both belonging to the device's vendor and to third parties). This has been typically done by relying on device modification or logging capabilities of specific platforms~\cite{giese2021amazon,celik2018sensitive,wang2018fear,jia2017contexiot}.
Other work used black-box experiments to identify how the viewing habits gathered by smart TVs~\cite{maheshwari_2018,ft19iot} and smart speakers interactions~\cite{iqbal-imc23} were shared with advertisers.
Finally, other studies that focus on web~\cite{bashir-pets-2018} and mobile apps~\cite{pan-pets-2018} have relied on instrumentation~\cite{reardon201950,han2020price}, automation, and network traffic interception~\cite{ren2018appversions,recon:mobisys} not available in the context of voice assistants.
In this work, we specifically address voice assistant, which are cloud services accessed through smart speakers or apps, and not devices themselves. Moreover, profile inference typically happens in the cloud and not the device itself. 
For this reason, we cannot use device modification or network traffic analysis as done in prior related work.
Similarly to some of the work above, we use a black-box approach, where we only access entry points (i.e., our voice queries) and exit points (i.e., user labels) from data disclosure requests mandated by law. The main differences is that we do not rely on third-parties such as advertisers, since we target first-party (i.e., vendor) labels and not how such labels are shared or used.

\section{Context, Definitions, and Goals}

This section describes the context, the definitions, and the goals of this work.

\subsection{Context}

The integration of voice assistants into daily life has revolutionized how we interact with technology, offering convenience and efficiency in performing a wide array of tasks. However, this interaction brings significant privacy implications, particularly concerning the profiling capabilities of these systems. Profiling by voice assistants involves analyzing voice commands, searches, and interactions to construct detailed profiles of users~\cite{google-privacy-policy,amazon-privacy-policy,apple-advertising}. These profiles can reveal private personal information, including demographic details, that users might prefer to keep confidential~\cite{google-myadcenter,amazon-adprefs,apple-privacy}.


The implications of such unintentional data sharing are multifaceted. Firstly, there is the risk of unauthorized access to sensitive information, where personal data could be exposed to third-parties or inadvertently leaked through security breaches~\cite{alexa_data_shared_with_stranger,smart_speaker_eavesdropping,googlehomeminiflaw17}. Secondly, the demographic and personal information obtained from interactions can be used to tailor marketing efforts in a highly targeted, and potentially intrusive, manner~\cite{iqbal-imc23}. 

Even though users, thanks to privacy laws~\cite{gdpr,ccpa}, can now access their profile and even opt out~\cite{google-myadcenter,amazon-adprefs,apple-privacy}, the opaque nature of data processing and profiling practices by voice assistant providers complicates users' ability to control what information is collected about them and how it is used~\cite{gunawan-2023-chi}. Many users are unaware of the extent to which their data contributes to their digital profile or the ways in which this profile influences their interactions with technology and content online~\cite{tabassum2023exploring,lau2018alexa,malkin2019privacy}.

Understanding the privacy implications of interacting with voice assistants is important for recognizing the potential risks to personal privacy. It highlights the importance of transparent data practices, robust privacy settings, and the need for users to be informed and cautious about the information they share through these increasingly ubiquitous services. 


\subsection{Definitions}
\label{sec:assumptions_and_definitions}
In the context of our study on the privacy implications and user profiling capabilities of voice assistant platforms, we refine our definitions to ensure clarity and precision in our analysis.

\vspace{-0.09in}
\paragraph{Voice assistants}
We define voice assistants as software components designed to interpret and execute voice commands, enabling them to perform various tasks. 
Voice assistants can receive voice inputs from users and provide audible feedback or take actions based on the analysis of these inputs.
The voice assistants we consider in this study, selected based on their popularity in the US ~\cite{statista-va}, are the following: Google Assistant~\cite{googleassistant}, Alexa~\cite{alexa}, and Siri~\cite{siri}. All these voice assistants are cloud-based, meaning that the user-controlled devices used to access them (e.g., smart speakers) simply forward the recordings to them, with the processing (and profiling) happening in the cloud.
This process is typically detailed in the privacy policies pages of each voice assistant.

\vspace{-0.07in}
\paragraph{Profiling labels}
We define profiling labels as the key components within a user profile, derived from user data and interactions with the voice assistants. These labels demonstrate specific characteristics that distinguish users from one another. We categorized them into two types: demographic and inference labels. Demographic labels capture the user's demographic attributes, such as relationship status, educational background, etc. These labels can either be based on user explicit data or be derived from user’s interactions. In contrast, inference labels offer insights into the user's preferences and interests in products or ad categories. These labels usually are inferred from the user’s interactions. Both types of labels enable personalized experiences and targeted content delivery.

\vspace{-0.08in}
\paragraph{Persona} 
A key aspect of our experimental methodology is to train voice assistant user accounts as a certain \emph{persona}.
Taking inspiration from previous work~\cite{barford-www-2014,bashir-2016-usenix,farooqi2020canarytrap,cook2020inferring,bashir2016tracing,carrascosa15conext,lecuyer2015sunlight,agarwal2020www}, we conceptualize a persona as a composite profile consisting of one or more specific profiling labels that collectively define the perceived characteristics, preferences, or demographic details of a user as interpreted by a voice assistant.
To construct a specific persona with targeted labels, we curate a series of tailored queries and ask them to the voice assistant. For instance, one might create a persona in Google's ecosystem that is identified as ``Single'' or ``Renter'', or in Amazon's ecosystem as having an interest in ``Fashion''. The definition of a persona is contingent upon the profiling mechanisms of each voice assistant, which use interaction data to assign labels according to their user characterization methods.

\vspace{-0.08in}
\paragraph{Training queries}
We define training queries as a curated set of queries intentionally designed to shape or guide the development of a targeted persona. The purpose of these queries is to direct the profiling algorithms of the voice assistant towards converging on a specific label that accurately reflects the intended persona. 

\vspace{-0.08in}
\paragraph{Modalities of interaction}
We define modalities of interaction as the various interaction methods through which users can access information from each voice assistant company. These interactions can occur through voice inputs via voice assistant interfaces or through text-based queries on web interfaces. Note that we do not distinguish modalities between vocal voice assistant interactions and textual non-web voice assistant interactions using accessible interfaces designed for people who cannot speak, since during our preliminary experiments we observed that such interactions are processed in the same way and produce the same results.

\subsection{Goals}
\label{sec:research_questions}
The goal of this study is to explore if and how voice assistants profile their users. Specifically, we will address the following research questions (RQ):
\noindent \textit{RQ1. Before experimentation, does the voice assistant have any prepopulated labels?} \\  
    This research question aims to determine if the system begins with any predefined labels on newly created accounts before any interaction with the system.
    Having wrong initial labels may affect user experience (e.g., irrelevant recommendations) and mislead advertisers, with the potential of wrong labels being propagated to third parties.
    We address this question by creating 200 accounts, and then waiting for labels to appear.
    
\noindent \textit{RQ2. After experimentation, does profiling actually happen and to what extent?} \\  
    We aim to quantify the extent of user profiling by voice assistants in terms of accuracy and consistency, assessing the depth and breadth of data collection. The purpose is to improve transparency over profiling practices.
    To answer this question, we perform 460 persona-based voice interaction large-scale experiments using 19 different personas for Google, 7 for Amazon, and 5 for Apple (see Section \ref{sec:Methodology} for a detailed description of our experiments and justification for the selected number of personas per platform).


\noindent \textit{RQ3. Does the interaction modality affect profiling?} \\
    This question investigates whether different modalities of interaction (e.g., web vs. voice) have distinct impacts on the scope and accuracy of user profiling.
    This is important to understand whether users are profiled (or misprofiled) differently when accessing information through a voice assistant with respect to another modality such as the web.
    To achieve this goal, we perform web persona-based experiments performing the same queries, but with a different modality, and then compare the results with our voice results.

\noindent \textit{RQ4. How to mitigate voice assistant profiling?} \\
    This research question focuses on identifying whether the current strategies to mitigate the privacy risks associated with user profiling by voice assistants are sufficient enough to protect users from being misprofiled or from being profiled against their will.
    To answer it we investigate whether we can effectively remove or change profiling labels from our persona accounts and the level of granularity (e.g., partial  vs. complete opt out).
In the next sections, we detail the methodology employed to investigate these RQs (Section~\ref{sec:Methodology}), followed by an analysis of the findings (Section~\ref{sec:Experimental_Results_of_Profiling_Activities}), and a discussion on the implications of our results (Section~\ref{sec:discussion}).

\subsection{Ethics}
\label{sec:ethics}

This research project involves only the authors of the paper and their direct collaborators listed in the acknowledgment section; no external research subjects were recruited.
The experiments were meticulously performed one at a time to minimize any disruption or disturbance to the voice assistants being tested. 

We did not need to disclose our findings with the companies hosting the voice assistants since their privacy policies already acknowledge the profiling activities associated with their services. Instead, our focus has been to provide greater transparency regarding these profiling activities and their implications for privacy.
\section{Methodology}
\label{sec:Methodology}

\begin{figure}[t]
  \centering
  \includegraphics[width=1.03\linewidth]{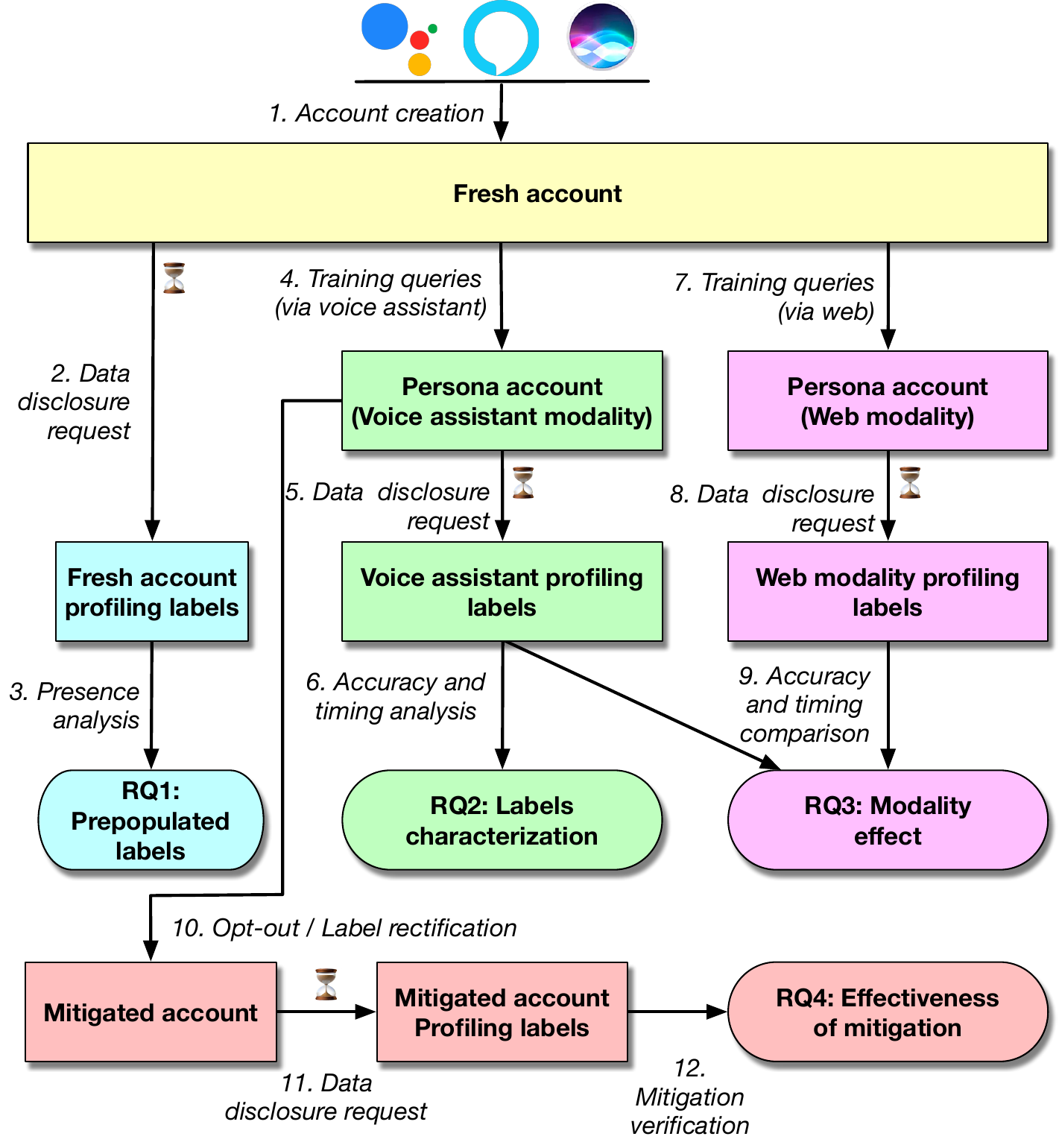}
  \caption{Methodology overview. Step 1 is the creation of fresh accounts. We perform Steps 2--3 to determine prepopulated labels before experimentation (RQ1). We perform Steps 4--6 to characterize profiling after experimentation (RQ2). We perform Steps 7--9 to analyze the effect of the modality (RQ3). Finally, we perform Steps 10--12 to evaluate profiling mitigation (RQ4). Steps marked with ``$\hourglass$'' include a waiting period to allow sufficient time for profiling to occur.}
  \label{fig:methodology}
\end{figure}

In this section we describe the methodology we used for answering our research questions.
Figure~\ref{fig:methodology} shows a summary of our methodology, divided into steps.
In the first step we prepare the voice assistant for our experimentation, i.e., we create fresh accounts and reset the smart devices we use to access them.
We then perform Steps 2--3 to understand prepopulated labels (RQ1); in particular, in Step 2 we perform a data disclosure request to obtain such prepopulated labels, while in Step 3 we analyze the presence of such labels.
In Steps 4--6 we turn the fresh accounts into profiled accounts using persona-based voice interactions and then analyze their profiling labels (RQ2). Specifically, in Step 4 we perform training voice queries for different personas; in Step 5 we obtain profiling labels following a data disclosure request; and in Step 6 we analyze such labels, checking their accuracy and time needed to profile.
Steps 7--9 are similar to 4--6, but using the web modality instead of the voice modality, so that we can see whether the modality has an effect on profiling (RQ3).
Finally, we perform Steps 10--12 to understand the effectiveness of the profiling mitigation tools offered by each voice assistant (RQ4). Specifically, in Step 10 we use opt-out and label rectification mechanisms to remove or amend the labels; in Step 11 we perform a data disclosure request for obtaining the new mitigated labels (after re-enabling profiling and waiting 10 days); and in Step 12 we verify that profiling mitigation actually worked (i.e., ensuring that labels are still absent or amended).
Note that before proceeding with the data disclosure request steps, we allow some time for profiling to occur, as profiling is not instantaneous.

In the next paragraphs, we introduce our testing environment (Section~\ref{sec:testbed_description}), our platform-independent approach for preparing for the experiments and ensuring minimal account contamination (Section~\ref{sec:interaction_method_overview}), 
our platform-agnostic experimental approach (Section~\ref{sub:unified-methodological-framework}),
and the platform-specific experimental approach for Google assistant (Section~\ref{sec:goole_experimens_manual}), Amazon Alexa (Section~\ref{sec:alexa_experimens_manual}), and Apple Siri (Section~\ref{sec:siri_experiments_manual}).
We then conclude this section with a brief discussion of early experiments (Section~\ref{sub:preliminary-experiments}) and with a summary and timeline of all our experiments per RQ (Section~\ref{sub:experiments_summary}). Note that during the 20 months of experimentation we conducted three major batches of experiments, the first in Spring 2023, the second in Fall 2024, and the third in Summer 2024, details of which are shown in Table \ref{tab:experiments-timeline-accounts}. For the rest of the paper we collectively refer to the first two early batches as preliminary experiments, and to the third most-recent one as large-scale experiment.

\subsection{Testing Environment}
\label{sec:testbed_description}
Our testing environment for evaluating voice assistant devices comprises three main components: (\emph{i}) the voice assistant platforms being tested; (\emph{ii}) a PC equipped with a Python script to automate the playback of audio recordings; and (\emph{iii}) a pair of non-smart speakers connected to the PC, tasked with playing the recordings to the voice assistant. 

To access Google, Alexa, and Siri voice assistants, we used both smart speakers and smart phones/tablets since during preliminary testing we noticed that the devices used to access the voice assistants have no impact with respect to profiling. Specifically, we use the following voice-assistant enabled devices:

\begin{itemize}
    \item \textbf{Google Assistant Access:} 
        (i) 3 iPhone 7 devices running iOS 15.8.3,
        (ii) 1 iPhone 8 running iOS 16.7.8, and
        (iii) 1 iPhone 13 running iOS 16.0.
    The Google Home app version on all iPhones was 1.9.115.901.

    \item \textbf{Alexa Access:} 
        (i) 1 Amazon Echo Dot (1st generation),
        (ii) 3 Amazon Echo Dot (3rd generation), and
        (iii) 1 Amazon Echo Dot (5th generation).
    The Alexa Home app was installed on an iPhone 13 Pro running iOS 17.5.1, with Alexa app version 2024.15.

    \item \textbf{Siri Access:} 
        1 iPhone 13 Pro Max running iOS 16.0.
\end{itemize}

For each voice assistant, we created distinct personas, each associated with a persona-specific set of English training queries. To guarantee the consistency of our experiments across multiple accounts of a same persona, we pre-recorded the queries we intended to ask to voice assistants using the voice of one of the authors.
Then, we developed an automated Python script to systematically play back these recordings. The process involves the script first triggering the voice assistant device's wake word (e.g., ``Hey Siri''), followed by a single query. It then pauses for a one-minute interval, which during preliminary testing proved to be a sufficient window for all our voice assistants to process and respond to the query. Upon completion of this interval, the script proceeds to play the next query.
During the whole experimentation, we manually supervised the automatic execution of the experiments to ensure that the voice assistant under test always activates (i.e., it recognizes the wake word) and always understands the voice query correctly (i.e., its answers are relevant to the questions).
This methodical approach ensures our experiments are repeatable under identical conditions.




\subsection{Experiment Preparation}
\label{sec:interaction_method_overview}
In conducting our experiments to understand voice assistant profiling, we adopted a rigorous method to maintain the independence and isolation of each device and account. This was imperative to ensure that our results accurately reflected the profiling behavior of voice assistants without the influence of external variables.

For each voice assistant platform, we created several fresh accounts (one for each persona instantiation)  using the Chrome web browser in private mode to minimize the chances of being tracked.

Since the account creation process required a phone number for SMS verification, we obtained a new phone number for each account to ensure independence. Each phone number was used only once per platform, ensuring that no two accounts shared the same phone number. By using a browser in private mode and by not reusing phone numbers across accounts within the same platform, we minimize any potential cross-contamination of data or profiling influences that could compromise the integrity of the experiment.

Additionally, we reset the device used to factory settings to access the voice assistant before logging in with a fresh account.
%
%
This ensured that no residual data from the previous account could influence the voice assistant's behavior.

Finally, to prevent our IP address from affecting the profiling process, we connected the smart devices used to access voice assistants through a VPN using the same IP, thus
ensuring that any observed differences in profiling were attributable to the interactions specific to each account and not to external network factors.

\subsection{Common Experimental Approach}
\label{sub:unified-methodological-framework}

We employed a common experimental approach across all the voice assistants we tested, as explained below. The following subsections highlight the specific variations in our methodology tailored to each voice assistant.

\vspace{-0.08in}
\subsubsection{Profiling and Data Access}
Our study began with determining which profiling labels are observable and how they can be accessed on each platform. Since each platform employs different approaches to categorize user information, we systematically searched for accessible profiling labels on each, exploring their data access tools and documentation to prepare for evaluating their profiling behaviors.

\vspace{-0.08in}
\subsubsection{Persona Choices}
After identifying the accessible profiling labels of each system, we designed personas, each targeting a specific profiling label. This strategy enabled us to develop tailored queries that guide the system towards recognizing each profiling label, and allowed us to study the system's behavior for each label.
For each persona, we created 10 accounts per persona to ensure a meaningful sample size while efficiently managing computational resources and time constraints. Previous studies in similar contexts have shown that 10 samples per persona are sufficient to detect consistent patterns and trends.

\vspace{-0.08in}
\subsubsection{Query Choices}
To develop the queries for our experiments, we employed prompt engineering with the goal of generating realistic and relevant questions that align with typical user interactions and engage with each voice assistant's profiling system.
We first used ChatGPT-4o~\cite{chatgpt4o} to assist in this process, ensuring that the queries were not influenced by the biases of the authors. 
We then carefully reviewed and manually refined the queries based on consensus among the authors.  
The tailored prompts used for this purpose are included in Appendix ~\ref{sec:query-generation-prompts}.

\vspace{-0.07in}
\subsubsection{Waiting Periods} 
Based on our preliminary experiments, we implemented an initial "prepopulation" one-week waiting period to allow any prepopulated labels to appear, as we observed that these labels typically emerge within this timeframe. This approach also enabled us to identify any pre-existing labels, ensuring that we excluded accounts where the target label was already present from the corresponding experiments.

After this initial waiting period, we deployed a set of training queries to trigger account profiling. Profiling by the voice assistants is not immediate either, and each platform requires a "profiling" waiting period before profiling labels start becoming observable. We determined this waiting period experimentally, and chose a standardized one-week waiting period for all platforms. This decision was informed by our preliminary experiments conducted in Spring and Fall of 2023 (see Section~\ref{sub:preliminary-experiments}), which indicated that profiling labels typically emerge within this timeframe (by 5.9 days for Google, and by 5.5 days for Alexa). 

In our most recent, large-scale experiments, we observed that profiling happens slower than in our preliminary experiments (see Section 
\ref{sec:Experimental_Results_of_Profiling_Activities} and more specifically Figure \ref{fig:google-time-comparison}).
Motivated by this, 
if some target labels were not achieved within the initial one-week period, we generated a new set of queries using the prompts described in Appendix~\ref{sec:query-generation-prompts} and waited for a new one-week period before re-observing the profiles. We continued this monitoring process for a total of four weeks, at which point we observed that profiles were stabilized.
(This approach additionally helped maintain consistent interaction levels, reducing the risk of accounts being flagged due to insufficient activity.)




\vspace{-0.08in}
\subsubsection{Modality Experiments}
To assess the impact of different interaction modes on user profiling, we conducted experiments using two primary querying methods: web searches and voice commands. For each service, we replicated persona experiments in a web browser environment, logged into new accounts, and compared the results to those obtained via voice interactions. This approach allowed us to explore how each modality influences the creation and updating of user profiles.

\vspace{-0.08in}
\subsubsection{Mitigation Experiments}
We conducted mitigation experiments 
beginning with accounts associated with personas that had successfully reached their target profiling labels. 
Once these accounts have been profiled, we opted out the accounts to stop data collection and profiling. Later, we opted the accounts back in and monitored their profiles over time to see if opting out had a lasting impact on the system's ability to profile based on past interactions.

\smallskip
In the rest of the paper, for experiment counting purposes, we count account actions, such as checking for prepopulated labels in an account (RQ1), asking a batch of queries to an account (RQ2, RQ3), and opting out of profiling for an account (RQ4), as one experiment each.

\subsection{Google Assistant Experiments}
\label{sec:goole_experimens_manual}

\subsubsection{Profiling and Data Access}
Google's ad personalization strategy utilizes comprehensive user interaction data from its services and it enables users to modify their advertisement preferences in their Google Account settings~\cite{google-privacy-policy}. Google categorizes this information into multiple demographic categories, outlined as follows:

\begin{itemize}
    \item \emph{Relationship:} whether the user is single, married, etc.
    \item \emph{Household Income:} estimated income bracket.
    \item \emph{Education:} highest level of education achieved.
    \item \emph{Industry:} sector of employment.
    \item \emph{Employer Size:} rough size of the user's place of employment.
    \item \emph{Homeownership:} whether the user owns or rents their home.
    \item \emph{Parenting:} whether the user has children.
\end{itemize}

Each of the above demographic categories has several possible options referred to as profiling labels. Google allows users to view their profiling labels in real-time through its My Ad Center interface~\cite{google-myadcenter}, encompassing services like Google Assistant.

In addition to demographic categories, Google also profiles users based on interest labels (as defined in Section \ref{sec:assumptions_and_definitions}). We monitored the impact of interactions on the generation of these interest labels by systematically reviewing the My Ad Center interface~\cite{google-myadcenter} over a 4-week period. However, we did not observe any evidence of interest labels being assigned through interactions with Google Assistant, suggesting that Google likely does not derive interest labels from voice interactions. Consequently, we focused our analysis of Google profiling on demographic labels, as they were the only ones demonstrably affected by voice interactions.

\vspace{-0.08in}
\subsubsection{Persona Choices}
Given Google's demographic-based profiling system, our selection process for personas included all possible demographic labels to ensure comprehensive targeting \footnote{\emph{Industry} and \emph{Employer Size} demographic categories were excluded due to their generalized nature, which complicates precise targeting.}. We considered each label as a potential persona. This allowed us to create a set of 19 distinct personas for Google profiling experiments.

\vspace{-0.08in}
\subsubsection{Query Choices}
For Google-specific experiments, we tailored prompts specifically for Google Assistant to align with Google's profiling system. The prompt used for this purpose is included in Appendix~\ref{sec:google-query-generation-prompt}.
For each persona (profiling label), ten distinct Google account samples were created. 
We designed 20 queries for each profiling label. This number was chosen as it was sufficient to trigger the target profiling labels based on insights from preliminary experiments. This allowed us to systematically evaluate the impact of query type on Google’s profiling accuracy.

\vspace{-0.08in}
\subsubsection{Mitigation Experiments}
In addition to opting-out, we also investigated the ability to selectively amend and delete labels offered by Google. In this case, our methodology does not include a waiting period since label deletion and editing are processed immediately (i.e., the change is immediately visible in Google My Ad Centre~\cite{google-myadcenter}).

\subsection{Amazon Alexa Experiments}
\label{sec:alexa_experimens_manual}
\subsubsection{Profiling and Data Access}
On Amazon's website~\cite{amazon}, there is no explicit way for users to visualize their personal data categorization, except for explicitly self-assigned labels available through the Amazon Profile Hub~\cite{amazon-profile-hub}. (The Amazon Profile Hub allows users to manually enter their demographic and interest labels, but it does not automatically infer these labels based on user behavior.)
To access inferred profiling data, Amazon facilitates Data Subject Access Requests (DSAR)~\cite{amazon-request-data}, allowing users to obtain specific data collected by Amazon.

From preliminary testing, we found out that information about user profiling is mainly found in the ``Advertising'' category, highlighting where users can find information related to how they are profiled for advertising purposes.
After downloading information from the Advertising category, users are given the following files:
%
    (i) \emph{Advertising.OptOut.csv} which tracks user opt-outs of profiling by country,
    (ii) \emph{Advertising.AmazonAudiences.csv} which lists categories of products that interest the user 
    according to Amazon's analysis,
    (iii) \emph{Advertising.3PAudience.csv} which identifies user interests based on third-party data,
    (iv) \emph{Advertising.AdvertiserAudiences.csv} which details which advertisers consider the user part of their target audience, and
    (v) \emph{Advertising.AdvertiserClicks.csv} which records instances where the user clicked on ads.

Upon interaction with an Amazon profile, the ``AmazonAudiences'' file reveals specific interest labels, categorizing the user's interests. These labels provide the categories of content and products that align with the user's interests.
We did not find evidence of demographic labels except for gender, only when explicitly declared by the user.
For this reason, when analyzing Amazon, we focus on interest-based labels that are directly tied to user behavior and product engagement.

\vspace{-0.08in}
\subsubsection{Persona Choices}
Given the fundamental difference between the labeling systems of Google and Amazon—demographic vs. interest-based—it was necessary to select personas that align with Amazon’s focus on specific user interests rather than broader demographic categories. Our selection was based on the maximum number of relevant interest labels we could identify from the available data, providing a solid foundation for a robust experimental analysis. This approach ensures that the personas are tailored to the nuances of Amazon’s profiling system.
\\
Amazon does not disclose the full range of interest labels it assigns to user profiles. To find interest labels to target for our experimentation, we performed Data Subject Access Requests (DSAR) for all the authors involved in this study. Upon receiving the data, we analyzed the ``AmazonAudiences'' file in each author's profile, where we identified a range of interest labels. We selected all In-Market first-level interest labels that appeared across the profiles, ensuring that our chosen personas reflected a broad spectrum of interests within Amazon’s system.

The seven personas derived from this process are as follows:
\begin{itemize}
    \item \textbf{Video Entertainment Persona:} Targeted at users interested in movies and television shows.
    \item \textbf{Fashion Persona:} Focused on users passionate about fashion.
    \item \textbf{Beauty and Personal Care Persona:} Designed for users interested in beauty and personal care products.
    \item \textbf{Electronics Persona:} For users with an interest in electronics.
    \item \textbf{Toys and Games Persona:} Targeted at users interested in toys and games.
    \item \textbf{Pet Supplies Persona:} Designed for users who frequently inquire about pet products.
    \item \textbf{Books and Magazines Persona:} Focused on users interested in books and magazines
\end{itemize}

\vspace{-0.08in}
\subsubsection{Query Choices}
For Amazon, we customized prompts for its voice assistant to align with their profiling system. Prompt details are in Appendix~\ref{sec:amazon-query-generation-prompt}.
Since Amazon Alexa supports command queries related to e-commerce, unlike Google Assistant, we structured our queries into two distinct types to analyze their effects on profiling:
    (i) \textit{General Queries} from which users seek factual information that Alexa answers without any additional action, and
    (ii) \textit{Command Queries} which require Alexa to perform specific tasks, such as adding items to a shopping list or making a purchase. 
%
This categorization helps us examine how each type of query affects profiling on Amazon, highlighting its e-commerce strengths. For Google Assistant, which does not provide command queries related to demographics like Amazon, we did not split the queries, focusing instead on general demographic profiling.

For each persona, ten distinct Amazon account samples were created. For each query type we designed 30 queries. This number was chosen because, based on insights from preliminary experiments, 30 queries were sufficient to trigger the target profiling labels on Amazon. In contrast, we used 20 queries for the Google experiments, as this number was found to be optimal for triggering profiling labels in that context. The differences in query numbers were tailored to the specific characteristics of each platform. 




\vspace{-0.08in}
\subsubsection{Modality Experiments}
\label{alexa-modalities-of-interaction}
Since users can interact with Amazon services through voice and web interfaces, we compare these two modalities to verify how differently they affect profiling. 
However, we limit the comparison to command queries because the Amazon web interface primarily searches for similar products and allows actions such as adding items to the cart or shopping list, but does not support general queries as the voice interface does.

\vspace{-0.08in}
\subsubsection{Mitigation Experiments}
Unlike Google, Amazon does not provide a direct method to modify or selectively delete its interest labels. Instead, it offers an indirect approach through the ``Improve My Recommendation’’ page~\cite{amazon-iyr}. However, since this page is intended for enhancing recommendations and does not allow for direct visualization, removal, or editing of interest labels, we do not take it into account in this study.


\subsection{Apple Siri Experiments}
\label{sec:siri_experiments_manual}
\subsubsection{Profiling and Data Access}
Our research methodology included submitting Data Subject Access Requests (DSARs) through Apple's designated platform~\cite{apple-privacy}. 
Despite our efforts, the data retrieved from these requests did not reveal explicit profiling information related to Siri interactions neither during preliminary experiments nor by inspecting the personal Apple accounts of the authors.
For this reason, our experiment methodology could not include persona experiments targeting Apple Siri labels.
Instead, since Siri uses third-party services such as Google Search or Bing Search to answer generic voice queries, we perform indirect persona experiments targeting such third-party services.

To address RQ2, which involves characterizing profiling activity, we focused on a third-party entity responsible for generating search results for Siri. We hypothesized that this third party may also conduct profiling activities on behalf of Apple. 
Since we already had an established methodology for Google profiling, we considered the integration with Google search and linked newly created Google Accounts to the device used to access Siri.

\vspace{-0.08in}
\subsubsection{Persona and Query Choices}
For the indirect experiments targeting Siri, we selected the five personas from the Google experiments that demonstrated the highest accuracy. These personas were characterized by the labels: ``Single,'' ``Married,'' ``Renters,'' ``Advanced Degree,'' and ``In a Relationship.'' The decision to focus on these five personas was informed by the results from the Google experiments, described in Section \ref{sec:google_experimental_measurement_of_profiling_activities}, which indicated that the remaining personas had a low chance of being accurately profiled. We used the exact same queries designed for those personas in the Google experiments.


\vspace{-0.08in}
\subsubsection{Modality Experiments and Mitigation Experiments}
Since Apple does not offer a web modality to access Siri information, we could not perform modality experiments for Apple Siri. Regarding mitigation experiments, Apple offers an opt-out mechanism similar to the other two platforms; however, we could not test it since Siri did not produce profiling labels.

\subsection{Preliminary Experiments}
\label{sub:preliminary-experiments}
The previous discussion on our platform-specific experimental approach is based on our most-recent, large-scale experiments. Here, we briefly discuss some variations between these and our preliminary experiments. For a detailed presentation of these preliminary experiments, see Appendix~\ref{sec:previous-experiments-in-the-previous-paper-version}.
Most notably, the preliminary experiments were smaller in scale, for example, the Amazon experiments only studied two of the seven interest labels that we studied in our current experiments. As another example, the Google and Apple experiments involved a much smaller number of accounts, queries, and labels. Last, note that in our preliminary experiments we also studied both single-label and multi-label personas.
These preliminary experiments laid the groundwork for our final methodology, allowing us to make informed decisions on persona construction, query design, interaction modalities, and wait times.

\subsection{Experiments Summary and Timeline}
\label{sub:experiments_summary}

To summarize, we performed a total of \numexperiments{} experiments
and \numqueries{} training queries over a period of \nummonths{} months. The definition of experiment is mentioned in Section \ref{sub:unified-methodological-framework}.
An overview of all the experiments we performed and their dates and relationship with each research question is reported in Table~\ref{tab:experiments-timeline-accounts}, showing the number of labels, experiments per label, and queries per label, along with the timeline of execution. Note that for each experiment used to answer RQ1, RQ2, and RQ3 we used a fresh account per experiment, while for RQ4, we used previously profiled accounts.

\begin{table}[]
\small
\begin{tabular}{cccccc}
\hline
\textbf{VA}                       & \textbf{RQ}          & \textbf{Time} & \textbf{\# Labels} & \textbf{\# Exp.} & \textbf{\# Queries} \\ \hline
\multirow{11}{*}{\textbf{Google}} & \multirow{3}{*}{RQ1} & Spr 23        & -                  & 16               & -                   \\ \cline{3-6} 
                                  &                      & Fall 23       & -                  & 30               & -                   \\ \cline{3-6} 
                                  &                      & Sum 24        & -                  & 200              & -                   \\ \cline{2-6} 
                                  & \multirow{4}{*}{RQ2} & Spr 23        & 4                  & 32               & 20                  \\ \cline{3-6} 
                                  &                      & Fall 23       & 8                  & 16               & 20                  \\ \cline{3-6} 
                                  &                      & Win 24        & 8                  & 16               & 20                  \\ \cline{3-6} 
                                  &                      & Sum24         & 19                 & 190              & 60                  \\ \cline{2-6} 
                                  & \multirow{2}{*}{RQ3} & Spr 23        & 8                  & 16               & 20                  \\ \cline{3-6} 
                                  &                      & Sum 24        & 19                 & 95               & 60                  \\ \cline{2-6} 
                                  & \multirow{2}{*}{RQ4} & Win 24        & 8                  & 16               & -                   \\ \cline{3-6} 
                                  &                      & Sum 24        & 19                 & 95               & -                   \\ \hline
\multirow{9}{*}{\textbf{Amazon}}  & \multirow{2}{*}{RQ1} & Sum 23        & -                  & 16               & -                   \\ \cline{3-6} 
                                  &                      & Sum 24        & -                  & 70               & -                   \\ \cline{2-6} 
                                  & \multirow{3}{*}{RQ2} & Sum 23        & 2                  & 16               & 30                  \\ \cline{3-6} 
                                  &                      & Win 24        & 2                  & 12               & 30                  \\ \cline{3-6} 
                                  &                      & Sum 24        & 7                  & 70               & 30                  \\ \cline{2-6} 
                                  & \multirow{2}{*}{RQ3} & Win 24        & 2                  & 12               & 30                  \\ \cline{3-6} 
                                  &                      & Sum 24        & 7                  & 35               & 30                  \\ \cline{2-6} 
                                  & \multirow{2}{*}{RQ4} & Win 24        & 2                  & 12               & -                   \\ \cline{3-6} 
                                  &                      & Sum 24        & 7                  & 70               & -                   \\ \hline
\multirow{4}{*}{\textbf{Apple}}   & \multirow{2}{*}{RQ1} & Sum 23        & -                  & 12               & -                   \\ \cline{3-6} 
                                  &                      & Sum 24        & -                  & 50               & -                   \\ \cline{2-6} 
                                  & \multirow{2}{*}{RQ2} & Sum 23        & 4                  & 24               & 20                  \\ \cline{3-6} 
                                  &                      & Sum 24        & 5                  & 50               & 20                  \\ \hline
                                  &                      & \textbf{}     & \textbf{Total=}             & 1171              & 24530                \\ \cline{4-6} 
\end{tabular}
\caption{Overview of our experimental efforts, showing the number of labels, experiments, and queries per label for each research question (RQ) and voice assistant (VA). Experiments refer to number of account actions, see Section \ref{sub:unified-methodological-framework}. Total number of queries (with repetitions) and total number of experiments are also included.}
\label{tab:experiments-timeline-accounts}
\end{table}
\section{Results}
\label{sec:Experimental_Results_of_Profiling_Activities}

In this section we discuss the results of our experiments and the answers to our research questions. Unless otherwise stated, we are referring to the results of the most-recent, large-scale experiments.


\subsection{Prepopulated Labels in Profiles}
\label{sec:Exploring_Prepopulated_Labels_in_Profiles}

In this section we address RQ1: \emph{Before experimentation, does the voice assistant have any prepopulated labels?}

\subsubsection{Google}
\label{sec:google-prepopulated-labels-in-profiling}
We created 200 fresh Google accounts in summer 2024 and did not use them for one week to observe evidence of profiling labels prior to any user interaction, as mentioned in Section \ref{sub:unified-methodological-framework}. Our observations showed several pre-populated labels upon creation. As shown in Table~\ref{tab:final_prepopulated_labels}, the tags ``Homeowners'' and ``Not Parents'' were the most frequently assigned prepopulated labels, appearing in 195 out of 200 accounts. Other categories such as ``Relationship,'' ``Household Income,'' ``Education,'' ``Industry,'' and ``Employer Size'' also exhibited varying degrees of pre-populated information. For instance, 92 accounts were labeled ``In a Relationship,'' and 194 accounts were tagged with ``Moderately High Income.''

This phenomenon suggests an initial profiling stage where accounts are assigned characteristics that could influence subsequent personalization strategies.

\begin{table}[]
\small
\begin{tabular}{llr}
\hline
{\color[HTML]{333333} \textbf{Category}}                           & {\color[HTML]{333333} \textbf{Label}}          & \multicolumn{1}{l}{{\color[HTML]{333333} \textbf{Count/200}}} \\ \hline
{\color[HTML]{333333} }                                            & {\color[HTML]{333333} Not Enough Info}         & {\color[HTML]{333333} 104}                                    \\ \cline{2-3} 
{\color[HTML]{333333} }                                            & {\color[HTML]{333333} Married}                 & {\color[HTML]{333333} 4}                                      \\ \cline{2-3} 
{\color[HTML]{333333} }                                            & {\color[HTML]{333333} Single}                  & {\color[HTML]{333333} 0}                                      \\ \cline{2-3} 
\multirow{-4}{*}{{\color[HTML]{333333} \textbf{Relationship}}}     & {\color[HTML]{333333} In a Relationship}       & {\color[HTML]{333333} 92}                                     \\ \hline
{\color[HTML]{333333} }                                            & {\color[HTML]{333333} Not Enough Info}         & {\color[HTML]{333333} 5}                                      \\ \cline{2-3} 
{\color[HTML]{333333} }                                            & {\color[HTML]{333333} High Income}             & {\color[HTML]{333333} 1}                                      \\ \cline{2-3} 
{\color[HTML]{333333} }                                            & {\color[HTML]{333333} Moderately High Income}  & {\color[HTML]{333333} 194}                                    \\ \cline{2-3} 
\multirow{-4}{*}{{\color[HTML]{333333} \textbf{Household Income}}} & {\color[HTML]{333333} Average or Lower Income} & {\color[HTML]{333333} 0}                                      \\ \hline
{\color[HTML]{333333} }                                            & {\color[HTML]{333333} Not Enough Info}         & {\color[HTML]{333333} 196}                                    \\ \cline{2-3} 
{\color[HTML]{333333} }                                            & {\color[HTML]{333333} Highschool Diploma}      & {\color[HTML]{333333} 4}                                      \\ \cline{2-3} 
{\color[HTML]{333333} }                                            & {\color[HTML]{333333} Attending College}       & {\color[HTML]{333333} 0}                                      \\ \cline{2-3} 
{\color[HTML]{333333} }                                            & {\color[HTML]{333333} Bachelor's Degree}       & {\color[HTML]{333333} 0}                                      \\ \cline{2-3} 
\multirow{-5}{*}{{\color[HTML]{333333} \textbf{Education}}}        & {\color[HTML]{333333} Advanced Degree}         & {\color[HTML]{333333} 0}                                      \\ \hline
{\color[HTML]{333333} }                                            & {\color[HTML]{333333} Not Enough Info}         & {\color[HTML]{333333} 200}                                    \\ \cline{2-3} 
\multirow{-2}{*}{{\color[HTML]{333333} \textbf{Industry}}}         & {\color[HTML]{333333} Other}                   & {\color[HTML]{333333} 0}                                      \\ \hline
{\color[HTML]{333333} }                                            & {\color[HTML]{333333} Not Enough Info}         & {\color[HTML]{333333} 5}                                      \\ \cline{2-3} 
{\color[HTML]{333333} }                                            & {\color[HTML]{333333} Small Employer}          & {\color[HTML]{333333} 0}                                      \\ \cline{2-3} 
{\color[HTML]{333333} }                                            & {\color[HTML]{333333} Large Employer}          & {\color[HTML]{333333} 195}                                    \\ \cline{2-3} 
\multirow{-4}{*}{{\color[HTML]{333333} \textbf{Employer Size}}}    & {\color[HTML]{333333} Very Large Employer}     & {\color[HTML]{333333} 0}                                      \\ \hline
{\color[HTML]{333333} }                                            & {\color[HTML]{333333} Not Enough Info}         & {\color[HTML]{333333} 5}                                      \\ \cline{2-3} 
{\color[HTML]{333333} }                                            & {\color[HTML]{333333} Homeowners}              & {\color[HTML]{333333} 195}                                    \\ \cline{2-3} 
\multirow{-3}{*}{{\color[HTML]{333333} \textbf{Homeownership}}}    & {\color[HTML]{333333} Renters}                 & {\color[HTML]{333333} 0}                                      \\ \hline
{\color[HTML]{333333} }                                            & {\color[HTML]{333333} Not Enough Info}         & {\color[HTML]{333333} 5}                                      \\ \cline{2-3} 
{\color[HTML]{333333} }                                            & {\color[HTML]{333333} Not Parents}             & {\color[HTML]{333333} 195}                                    \\ \cline{2-3} 
\multirow{-3}{*}{{\color[HTML]{333333} \textbf{Parenting}}}        & {\color[HTML]{333333} Other}                   & {\color[HTML]{333333} 0}                                      \\ \hline
\end{tabular}
\caption{Distribution of demographic tags assigned to new Google accounts during an initial profiling phase. The table categorizes 200 accounts with different labels, with the frequency of each label presented.}
\label{tab:final_prepopulated_labels}
\end{table}

\vspace{-0.09in} 
\subsubsection{Alexa}
During all our experiments with fresh accounts, we have not seen preopopulated interest profiling labels.
The only change we have seen is the appearance of an empty Advertising.AmazonAudiences file after a few days, confirming the absence of any interest label.


\vspace{-0.09in}
\subsubsection{Siri}
We found that Apple Siri does not assign any profiling labels to users before any interaction, which shows there is no early stage of profiling activity in Apple upon account creation and prior to any user interaction.

\smallskip
\noindent \textit{Takeaway:} Google may assign prepopulated profiling demographic labels even before any user interaction occurs, while that does not happen for Alexa and Siri.

\subsection{Characterization of Profiling Activities}
\label{sec:Experimental_measurement_of_profiling_activities}

In this section we address RQ2: \emph{After experimentation, does profiling actually happen and to
what extent?}
For each voice assistant, we analyze the following profiling aspects: 
%
(i) \emph{Accuracy,} defined as the fraction of persona instantiations (i.e., accounts) 
that are correctly labeled after a given time period following a persona experiment, and
(ii) \emph{Time needed for profiling,} defined as the 
time required for the accounts to transition from an initial label to a different label. (In case of no change, we define this metric as ``N/A.'')


\vspace{-0.07in}
\subsubsection{Google}
\label{sec:google_experimental_measurement_of_profiling_activities}

In summer 2024, we conducted our experiments with Google to assess its profiling capabilities, as detailed in Section ~\ref{sec:goole_experimens_manual}. We used fresh Google accounts and began by waiting one week to allow any prepopulated labels to appear, as described in Section \ref{sub:unified-methodological-framework}. After this initial waiting period, we selected accounts that did not have the target profiling labels prepopulated and proceeded with the experiments by asking the queries.

To ensure a fair comparison across different labels, we excluded experiments for the ``Not parents,'' ``Moderately High Income,'' and ``Homeowners'' personas because these labels were consistently present across all accounts and we didn't want to count those as successful profiling outcomes.

\vspace{-0.1in}
\paragraph{Accuracy results.}
The accuracy of the accounts after a four-week monitoring interval time is shown in Table \ref{tab:google-voice-results}. The experimental findings for each profiling category are as follows:
%
    (i) \textit{Relationships Category:} Initially, no prepopulated labels existed for the relationship status of the accounts. Post-experiment, all labels achieved positive accuracy levels. The label ``Married'' stood out with the highest accuracy at 70\%, making it the most reliably identified status within this group. The other labels, such as ``Single'' and ``In a relationship,'' also displayed positive accuracy but were less accurate in comparison to the ``Married'' label, achieving 30\% and 10\% respectively. 
    (ii) \textit{Education Category:} In the Education category, all accounts were initially marked with ``Not enough info.'' As the study progressed, the label ``Advanced Degree'' showed an increase in accuracy, achieving a 50\% accuracy rate, marking it as the most accurately identified educational status in the experiment. In contrast, the other educational labels---``High school diploma,'' ``Attending college,'' and ``Bachelor’s degree''---remained at ``Not enough info'' throughout the monitoring period. 
    (iii) \textit{Homeownership Category:} All accounts were initially labeled as ``Homeowners.'' Despite this starting point, the ``Renters'' label was adjusted to an 80\% accuracy rate by the end of the experiment, demonstrating a noticeable correction from the initial label.
    (iv) \textit{Household Income and Parenting Categories:} These categories maintained strong label adherence. The initial labels ``Moderately high income'' and ``Not parents'' persisted without any changes throughout the duration of the study, indicating stability in these profiling categories.

The profiling results demonstrate distinct outcomes across various categories. In the ``Household Income'' and ``Parenting'' categories, the same prepopulated labels---``Moderately high income'' and ``Not parents''---were observed across all personas and remained unchanged throughout the experiments. Conversely, in the ``Homeownership'' category, adjustments were made to prepopulated labels, indicating changes based on profiling activities. Furthermore, all labels in the ``Relationships'' category achieved positive accuracy, showcasing effective profiling, while in the ``Education'' category, three personas maintained an accuracy of 0\%, indicating no effective profiling for these attributes.

These results confirm that profiling is actively occurring. The variability in profiling outcomes illustrate that, despite using a uniform methodology across all categories, the accuracy of the profiling results varies over different personas (profiling labels).

\noindent \textit{Takeaway:} Google profiles users based on their interactions and the profiling accuracy varies across different labels using the same methodology.


\begin{table}[]
\small
\begin{tabular}{llr}
\hline
\textbf{Profiling Category}                 & \textbf{Profiling Label}   & \textbf{Accuracy (\%)}                   \\ \hline
                                            & Married                    & 70\%                                     \\ \cline{2-3} 
                                            & Single                     & 30\%                                     \\ \cline{2-3} 
\multirow{-3}{*}{\textbf{Relationships}}    & In a relationship          & 10\%                                     \\ \hline
                                            & High Income                & 0\% (inaccurate)                         \\ \cline{2-3} 
                                            & Moderately high    & -                                      \\ \cline{2-3} 
\multirow{-3}{*}{\textbf{Household Income}} & Average or low    & 0\% (inaccurate)             \\ \hline
                                            & High school diploma        & 0\% (no tags)    \\ \cline{2-3} 
                                            & Attending college          & 0\% (no tags)    \\ \cline{2-3} 
                                            & Bachelor's degree          & 0\% (no tags)    \\ \cline{2-3} 
\multirow{-4}{*}{\textbf{Education}}        & Advanced degree            & 50\%                                     \\ \hline
                                            & Homeowners                 & -                                      \\ \cline{2-3} 
\multirow{-2}{*}{\textbf{Homeownership}}    & Renters                    & 80\%                                     \\ \hline
                                            & Not Parents                & -                                      \\ \cline{2-3} 
                                            & Infants         & 0\% (inaccurate) \\ \cline{2-3} 
                                            & toddlers        & 0\% (inaccurate) \\ \cline{2-3} 
                                            & preschoolers    & 0\% (inaccurate) \\ \cline{2-3} 
                                            & grade schoolers & 0\% (inaccurate) \\ \cline{2-3} 
\multirow{-6}{*}{\textbf{Parenting}}        & teenagers       & 0\% (inaccurate) \\ \hline
\end{tabular}
\caption{Accuracy of label assignment for Google experiments across different profiling labels (in percentage): ``no tags'' indicates all labels were categorized as ``Not Enough Info'', while ``inaccurate'' denotes incorrect label assignment.}
\label{tab:google-voice-results}
\end{table}

\begin{figure}[t]
  \centering
  \includegraphics[width=0.8\linewidth]{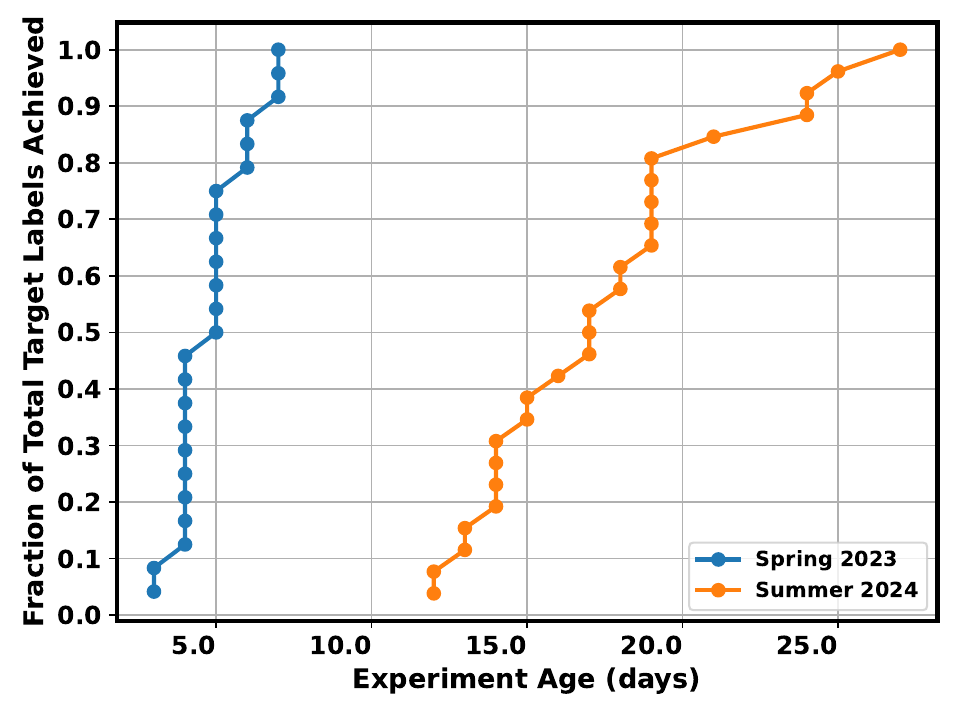}
  \caption{Comparison of target label achievement between Spring 2023 and Summer 2024 experiments. The Spring 2023 experiments show faster profiling progression, while the Summer 2024 experiments indicate a slower rate of achieving target labels.}
  \label{fig:google-time-comparison}
\end{figure}

\vspace{-0.05in}
\paragraph{Time needed for profiling.}
On average, it took 18.0 ± 4.1 days for all the profiling labels to be generated in our Summer 2024 experiments, based on Table \ref{tab:google-modalities}. This indicates that users should not expect to see the outcomes of their activity until this time has passed.

\vspace{-0.07in}
\paragraph{Comparison with Preliminary Experiments Results}
In Spring 2023, we conducted preliminary experiments focusing on the profiling accuracy of four labels: ``Married,'' ``Bachelor's degree,'' ``Single,'' and ``High school diploma.'' During that period, voice experiments resulted in an average accuracy of 80\% across these labels, with an average time to label change of 4.8 ± 1.1 days, as detailed in Appendix \ref{sec:previous-experiments-in-the-previous-paper-version}. In our final Summer 2024 experiments, there have been some changes. The average accuracy for voice experiments for these four labels has decreased to 20\%, while the average time to label change has increased to 15.9 ± 4.3 days for the same four labels, as shown in Table \ref{tab:google-voice-results}. Figure \ref{fig:google-time-comparison} further highlights this shift, showing a noticable delay in the timing of label convergence in the Summer 2024 experiments compared to Spring 2023.

\smallskip
\noindent \textit{Takeaway:} The accuracy and speed of profiling by Google voice assistant have decreased over time, with noticeable differences across the four labels tested, despite using the same methodology.

\subsubsection{Alexa}
\label{sec:alexa_experimental_measurement_of_profiling_activities}

\emph{Accuracy results.}
In our analysis of Alexa's profiling accuracy for different interest labels, we observed changes only in the ``AmazonAudiences'' file. The impact of the training questions on profiling accuracy is detailed in Table~\ref{tab:alexa-general-command}. The findings show that general questions do not results in profiling, while command-based queries result in 100\% profiling accuracy.
This results supports our initial hypothesis that for Amazon to assign relevant personas to accounts, the queries must be directly related to its product offerings rather than consisting of general inquiries.

\begin{table}[]
\small
\begin{tabular}{llcc}
\hline
\textbf{Query}                     & \textbf{Persona} & \multicolumn{1}{l}{\textbf{Acc (\%)}} & \multicolumn{1}{l}{\textbf{Avg Time (Days)}} \\ \hline
                                   & Fashion                                  & 0                                                                 & N/A                                                                   \\ \cline{2-4} 
                                   & Video                       & 0                                                                 & N/A                                                                   \\ \cline{2-4} 
                                   & Beauty                  & 0                                                                 & N/A                                                                   \\ \cline{2-4} 
                                   & Electronics                              & 0                                                                 & N/A                                                                   \\ \cline{2-4} 
                                   & Toys                            & 0                                                                 & N/A                                                                   \\ \cline{2-4} 
                                   & Pet Supplies                             & 0                                                                 & N/A                                                                   \\ \cline{2-4} 
\multirow{-7}{*}{\textbf{General}} & Books                       & 0                                                                 & N/A                                                                   \\ \hline
                                   & Fashion                                  & 100\%                                                        & 8.2 ± 1.0                                                             \\ \cline{2-4} 
                                   & Video                       & 100\%                                & 7.6 ± 1.4                                                             \\ \cline{2-4} 
                                   & Beauty                  & 100\%                                & 6.2 ± 1.5                                                             \\ \cline{2-4} 
                                   & Electronics                              & 100\%                                & 8.4 ± 1.4                                                             \\ \cline{2-4} 
                                   & Toys                            & 100\%                                & 7.6 ± 0.8                                                             \\ \cline{2-4} 
                                   & Pet Supplies                             & 100\%                                & 8.4 ± 0.5                                                             \\ \cline{2-4} 
\multirow{-7}{*}{\textbf{Commad}}  & Books                       & 100\%                                & 7.6 ± 1.0                                                             \\ \hline
\end{tabular}
\caption{Comparison of profiling accuracy and the mean time
required for accurate label convergence for Alexa’s general and command queries across different personas}
\label{tab:alexa-general-command}
\end{table}

\vspace{-0.08in}
\paragraph{Time needed for profiling.}
On average, it took Alexa 7.7 ± 1.3 days to generate profiling labels after our persona experiments (see Table~\ref{tab:alexa-general-command}). This indicates that, similarly to Google, users should not anticipate to see the outcomes of their activity until this time has passed.

\smallskip
\noindent \textit{Takeaway:} Users are profiled only based on command queries, such as adding items to a shopping cart or making a purchase, and such profiling is 100\% accurate.

\vspace{-0.07in}
\paragraph{Comparison with Preliminary Experiments Results}In both our preliminary and final experiments, command-based queries achieved 100\% profiling accuracy. However, the time required for accurate label convergence increased slightly, from 5.0 ± 0.82 days in the preliminary experiments (as shown in Table \ref{tab:alexa-general-command-old} in Appendix \ref{sec:previous-experiments-in-the-previous-paper-version}) to 7.7 ± 1.3 days in the final experiments (Table \ref{tab:alexa-general-command}). This suggests a slight variation in timing, though accuracy remains consistent.

\smallskip
\noindent \textit{Takeaway:} For the experimented personas, the accuracy of Amazon’s profiling through command-based queries has remained consistent over time, while the time required for profiling has slightly increased, indicating a slower response in the profiling process.

\subsubsection{Siri}
\label{sec:siri_experimental_measurement_of_profiling_activities}
\emph{Direct profiling.}
As discussed in Section~\ref{sec:privacy_polocies}, Apple's privacy policies indicate that they may infer user interests for targeted advertising~\cite{apple-advertising}.
However, our experimental submissions of Data Subject Access Requests (DSARs) for various persona-based Apple accounts yielded no profiling information. This outcome suggests that Apple does not profile users based on their Siri interactions.

\smallskip
\noindent \textit{Takeaway:} Despite Apple's policy statements allowing profiling, there is no evidence that Apple engages in profiling activities based on voice interactions through Siri.

\vspace{-0.07in}
\paragraph{Indirect profiling.}
We next investigated whether there is any profiling in the context of Apple Siri's integration with third-party search engines for answering user queries. 
Since we already had an established methodology for Google profiling, we considered the integration with Google search and linked newly created Google Accounts to the device used to access Siri. We used the queries for the top five selected personas of google voice experiments (see section \ref{sec:Experimental_measurement_of_profiling_activities}) to uncover any indirect profiling by Google, as evidenced by changes in the Google privacy dashboards linked to the Siri interactions.
Despite allowing a waiting period of four weeks after the completion of each of these experiments, we observed no new profiling activities in the Google accounts associated with the device used to perform the Siri queries. This lack of observed profiling activity suggests the absence of indirect profiling by Google for Siri voice searches.

\smallskip
\noindent \textit{Takeaway:} There is no evidence of indirect profiling by Google based on Siri voice queries, when Siri uses Google searches.

\subsection{Impact of Interaction Modality}
\label{sec:impact_of_interaction_modality_on_user_profiling}
This section addresses RQ3: \emph{Does the interaction modality affect profiling?}

To answer this research question, i.e., to evaluate the impact of various interaction modalities on user profiling practices, we systematically conducted experiments across different modalities for Google and Alexa, as detailed in Sections~\ref{sec:goole_experimens_manual}--\ref{sec:alexa_experimens_manual}. This approach allowed us to directly compare how distinct modes of interaction contribute to the profiling capabilities of the voice assistants.

Note that all the voice assistants we examined provide an option to input queries via typing, serving as an alternative for individuals who may be unable or unwilling to use voice queries. However, we do not categorize these textual interactions as distinct modalities. Unlike the web modality discussed in this section, these inputs are directly processed by the voice assistant.

\vspace{-0.08in}
\subsubsection{Google}
\label{sec:google_impact_of_interaction_modality_on_user_profiling}
In our investigation, we compared the profiling efficacy of voice and web modalities on Google's platform using the same persona queries detailed in Section~\ref{sec:goole_experimens_manual}. 

The results, summarized in Table~\ref{tab:google-modalities}, illustrate the differences in mean accuracy and the average time required for changes in Google's label profiling to be observed following the experiments with each modality. The web modality yielded a higher average accuracy of 62.86\%, compared to the voice modality, which exhibited a lower accuracy of 48\% across all labels. Furthermore, the average label convergence time for web modality was 2.2 ± 2.0 days, which was lower than the one for voice modality, which averaged 18.0 ± 4.1 days. These findings highlight superior performance of web modality in both accuracy and efficiency.

\begin{table}[]
\small
\begin{tabular}{llrc}
\hline
\textbf{Modality}                & \textbf{Persona} & \textbf{Acc (\%)}       & \multicolumn{1}{l}{\textbf{Avg Time (Days)}} \\ \hline
                                 & Married                  & 70\%                         & 16.1 ± 4.9                                   \\ \cline{2-4} 
                                 & Single                   & 30\%                         & 15.3 ± 2.6                                   \\ \cline{2-4} 
                                 & In a relationship        & 10\%                         & 25                                           \\ \cline{2-4} 
                                 & High school              & 0\%  & N/A                                          \\ \cline{2-4} 
                                 & Attending college        & 0\%  & N/A                                          \\ \cline{2-4} 
                                 & Advanced degree          & 50\%                         & 16.5 ± 1.5                                   \\ \cline{2-4} 
\multirow{-7}{*}{\textbf{Voice}} & Renters                  & 80\%                         & 19.6 ± 2.5                                   \\ \hline
                                 & Married                  & 80\%                         & 1.8 ± 1.3                                    \\ \cline{2-4} 
                                 & Single                   & 60\%                         & 1.0 ± 0.0                                    \\ \cline{2-4} 
                                 & In a relationship        & 60\%                         & 1.0 ± 0.0                                    \\ \cline{2-4} 
                                 & High school              & 20\% & 1.0 ± 0.0                                    \\ \cline{2-4} 
                                 & Attending college        & 60\% & 5.7 ± 0.5                                    \\ \cline{2-4} 
                                 & Advanced degree          & 60\%                         & 4.0 ± 2.2                                    \\ \cline{2-4} 
\multirow{-7}{*}{\textbf{Web}}   & Renters                  & 100\%                        & 1.0 ± 0.0                                    \\ \hline
\end{tabular}
\caption{Comparison of profiling accuracy and the mean time required for accurate label convergence across Google voice and web modalities. The labels that did not reach any level of accuracy (0\%) in both modalities were excluded in this table. }
\label{tab:google-modalities}
\end{table}
\smallskip
\noindent \textit{Takeaway:} Interactions with Google via web modality are more accurate and faster than those conducted through voice.

\vspace{-0.08in}
\subsubsection{Alexa}
\label{sec:alexa_impact_of_interaction_modality_on_user_profiling}

We explored the impact of different interaction modalities on user profiling by Amazon, focusing on voice and web queries as outlined in Section~\ref{sec:alexa_experimens_manual}. It is important to note, as mentioned in Section~\ref{sec:alexa_experimens_manual}, that we excluded web-based queries for the Video Persona due to the lack of a direct web interface equivalent for the voice command ``Add to Watchlist'' used for the Video Persona.

Table~\ref{tab:amazon-modalities} shows the web modality interaction results, where we used the same command queries using the web interface. (We could not use the web interface for general questions requiring information, since Amazon website does not provide such a service.) The web results show the same 100\% accuracy and an average time of 7.1 ± 1.4 days for labels to change in web experiments which is similar to the results of voice interactions (7.7 ± 1.3 days).

\smallskip
\noindent \textit{Takeaway:} For Amazon Alexa, the web modality of interaction profiles similar to the voice modality and with the same 100\% accuracy.

\begin{table}[]
\small
\begin{tabular}{cllc}
\hline
\textbf{Modality}               & \textbf{Persona}      & \textbf{Acc (\%)} & \multicolumn{1}{l}{\textbf{Avg Time (Days)}} \\ \hline
\multirow{7}{*}{\textbf{Voice}} & Fashion      & 100\%    & 8.2 ± 1.0                            \\ \cline{2-4} 
                       & Video        & 100\%    & 7.6 ± 1.4                            \\ \cline{2-4} 
                       & Beauty       & 100\%    & 6.2 ± 1.5                            \\ \cline{2-4} 
                       & Electronics  & 100\%    & 8.4 ± 1.4                            \\ \cline{2-4} 
                       & Toys         & 100\%    & 7.6 ± 0.8                            \\ \cline{2-4} 
                       & Pet Supplies & 100\%    & 8.4 ± 0.5                            \\ \cline{2-4} 
                       & Books        & 100\%    & 7.6 ± 1.0                            \\ \hline
\multirow{7}{*}{\textbf{Web}}   & Fashion      & 100\%    & 7.4 ± 0.5                            \\ \cline{2-4} 
                       & Video        & 100\%    & 8.6 ± 1.4                            \\ \cline{2-4} 
                       & Beauty       & 100\%    & 8.2 ± 1.2                            \\ \cline{2-4} 
                       & Electronics  & 100\%    & 6.4 ± 1.5                            \\ \cline{2-4} 
                       & Toys         & 100\%    & 6.4 ± 0.8                            \\ \cline{2-4} 
                       & Pet Supplies & 100\%    & 5.8 ± 0.7                            \\ \cline{2-4} 
                       & Books        & 100\%    & 7.0 ± 0.9                            \\ \hline
\end{tabular}
\caption{Comparison of profiling accuracy and the mean time required for accurate label convergence across Amazon voice and web modalities}
\label{tab:amazon-modalities}
\end{table}

\vspace{-0.06in}
\subsubsection{Siri}
\label{sec:siri_impact_of_interaction_modality_on_user_profiling}
Since Siri does not offer the option to search through a web interface, the way Google and Amazon do, we were unable to investigate how different modalities affected Siri profiling.

\subsection{Profiling Mitigation}
\label{sec:Mitigating_Privacy_Risks_in_Voice_Assistant_Profiling}
This section addresses our last research question (RQ4): \emph{How to mitigate voice assistant profiling?}


\vspace{-0.1in}
\subsubsection{Google}
\label{sec:google_Mitigating_Privacy_Risks_in_Voice_Assistant_Profiling}
Google provides users with two approaches to manage profiling via voice assistant interactions, tailored for those who prefer not to share their information for ad personalization, and those who prefer to have fine-grained control over it.

\paragraph{Direct modification of demographic labels.}
This feature allows users to personally adjust and confirm their demographic labels within the Google Privacy Dashboard~\cite{google-myadcenter}. By navigating to this site, individuals have the flexibility to review and modify their demographic information to align with their actual identifiers. This functionality empowers users to directly influence the profiling mechanism to better reflect their preferences.

After testing this feature, we confirmed that users can either change their existing demographic labels or confirm an existing tag of their choosing, ensuring that the ad content is closely aligned with their specified characteristics.

\vspace{-0.07in}
\paragraph{Selective opt-out from profiling.}
Google not only offers a general opt-out from profiling but also allows users to specifically disable profiling for certain categories ~\cite{google-myadcenter}. This ensures that users' interactions do not lead to the generation of new profiling labels within their account for those specific categories, thereby providing a tailored approach to privacy management.

To evaluate the impact of opting out on user profiles, we conducted an experiment with all accounts being profiled using web modality for all personas. These accounts demonstrated higher profiling accuracy than voice modality ones, as indicated in Table \ref{tab:google-modalities}. Following this, we opted out of profiling for these accounts. After a one-week break, we reactivated ad personalization and monitored the accounts daily for more than 10 days. 

Observations revealed that immediately after re-enabling ad personalization, all accounts were assigned the ``Not Enough Info'' label across all demographic fields. This condition remained unchanged after four weeks, evidencing that previous interactions prior to opting out did not influence subsequent profiling post-opting in. 

\smallskip
\noindent \textit{Takeaway:} Both mitigation mechanisms of Google (changing the demographic labels directly from the Google Privacy Dashboard and opting out from profiling) work as advertised and therefore are useful to mitigate mislabeling and unwanted profiling.

\vspace{-0.07in}
\subsubsection{Alexa}
\label{sec:alexa_Mitigating_Privacy_Risks_in_Voice_Assistant_Profiling}
To investigate whether Amazon has safeguards in place to enhance user privacy, we selected ``Do not show me interest-based ads provided by Amazon'' and deleted personal information from Amazon's ad systems, as found on Amazon's ad preferences page~\cite{amazon-adprefs}. After monitoring the accounts for four weeks, we found that the recommendations page showed generic ads, rather than personalized ads. A limitation of Amazon approach is that there is no selective way to remove or update the labels once they appear.

\vspace{0.02in}
\noindent \textit{Takeaway:} Also for the Amazon Alexa platform, choosing to opt out works as expected and is an effective way to limit profiling. However, Amazon does not provide tools to limit mislabeling by allowing users to selectively remove or rectify their labels.

\vspace{-0.06in}
\subsubsection{Siri}
\label{sec:siri_Mitigating_Privacy_Risks_in_Voice_Assistant_Profiling}

\noindent According to Apple's privacy policy~\cite{apple-advertising}, as mentioned in Section~\ref{sec:privacy_polocies}, users have the ability to deactivate Personalized Ads by adjusting their device settings, which prevents the delivery of ads tailored to their interests. However, we could not test this since, as discussed in Section~\ref{sec:siri_experimental_measurement_of_profiling_activities}, we did not observe any direct or indirect user profiling by Apple with respect to Siri interaction.

\smallskip
\noindent \textit{Takeaway:} Similar to other platforms, Apple provides tools to limit profiling in terms of opt-outs, but we could not test them because we never experienced profiling with Siri. For the same reason, we could not investigate the presence of tools for rectifying and selectively removing labels.

\subsection{Results from Preliminary Experiments}
\label{sec:preliminaryresults}

So far we have focused on reporting results from our most recent larger scale experiments. Here we summarize additional results from our preliminary experiments, see Appendix~\ref{sec:previous-experiments-in-the-previous-paper-version} for a detailed presentation. 

RQ1: Our experiments revealed that Google accounts may have prepopulated labels, whereas Amazon and Siri did not show any signs of prepopulated labels.

RQ2: Our experiments revealed that profiling does indeed happen.  Interestingly, we found that the less the number of labels per persona the faster and more accurate the labeling is. Specifically, while one- and two-label personas demonstrated good accuracy, four-label ones required repeated queries a few weeks after the first round of queries to show good accuracy. In our recent experiments, informed by this insights, we targeted one label across all personas. This approach allowed us to  cover all labels with more number of accounts per label and conduct large-scale experiments. 

RQ3: Our experiments revealed that Google web modality outperformed voice, while Amazon showed similar behavior between web and voice, both consistent with our recent findings. Note that since Siri does not offer the option to search through a web interface, we did not test the web modality for that as well.

RQ4: Our experiments revealed that mitigation tools are effective, consistently with our recent experiments. 
Specifically, for Google,
after opting out of profiling categories and reactivating ad personalization, we found that personas did not regenerate their previous labels, and, for Amazon, after disabling interest-based ads and deleting personal information, we observed a shift to generic recommendation ads. With Siri, we could not test the profiling mitigation tools due to not experiencing profiling with Siri.



\section{Discussion}
\label{sec:discussion}
This section describes privacy implications of our results, an overview of the systems' privacy policies, 
and limitations of our study.
\subsection{Implications of Our Findings}

\emph{Pre-populated labels.} In our experiments with Google Assistant, we observed that some newly created accounts had pre-populated labels, a phenomenon not observed with other platforms we studied, though the exact mechanisms by which these labels are generated remain unclear. It is possible that the platform is using indirect methods, such as IP address geolocation, to infer these labels~\cite{privacy-in-target-advertising}. Because these labels may not always be accurate, there is a risk that they could lead to displaying irrelevant personalized content to users or causing advertisers to target users based on potentially inaccurate data. It is worth noting that these prepopulated labels tend to be replaced as users begin interacting with the platform, as shown in Section~\ref{sec:google_experimental_measurement_of_profiling_activities}, indicating that Google may not have high confidence in the initial labels. 
Motivated by this, we believe it may improve transparency if confidence levels of labels are maintained and shared with users. 

\emph{Profiling accuracy.} 
On the platforms that exhibited profiling (i.e., Google Assistant and Alexa), we noticed a difference in how they handle profiling. Alexa's approach to profiling is deterministic, meaning it predicts user interests based on explicit information, resulting in 100\% accuracy and 100\% consistence, as shown in Table~\ref{tab:alexa-general-command}. 
However, Google's profiling tends to be probabilistic, meaning there is a chance of error depending on various factors like demographic labels, timing of experiments, and interaction modality, as shown in Tables~\ref{tab:google-voice-results} and~\ref{tab:google-modalities}. This introduces the risk of mislabeling, leading to similar concerns with prepopulated labels as discussed above. 
The lack of detailed information regarding how these systems operate and the undisclosed rationale behind label assignments make it hard to know the causes of this probabilistic behavior.

\emph{Profiling timing.}
On all platforms where profiling occurred, whether Amazon's or Google's approach, profiling typically began after user activity started, with the profiling process taking approximately 7.7 days on Amazon and 18.0 days on Google, as stated in Section \ref{sec:Experimental_measurement_of_profiling_activities}. During this time, a user's profile is being formed but not yet finalized. This presents a privacy concern: if a user requests their data too early (i.e., before the profiling period is complete), they may receive no information about profiling, leading them to believe they have not been profiled or have been profiled incorrectly. 
To promote transparency, we suggest that platforms disclose the profiling timeline and explore options to reduce and disclose the time required to complete a user's profile.

\emph{Modality.}
We have observed that Google exhibits varying profiling outcomes when the interaction modality changes, and both timing and accuracy are impacted depending on the label. This suggests that some platforms may employ different profiling methods between the web and voice assistant modality. Consequently, users of voice assistants face a higher privacy risk due to increased mislabeling and longer delays in profiling, as shown in Tables \ref{tab:google-modalities} and \ref{tab:amazon-modalities}. This finding aligns with prior research indicating that modality can influence the accuracy and effectiveness of user profiling~\cite{modality-influence}.

\emph{Mitigation.}
All platforms provided options to opt out from profiling, which helps address privacy concerns related to profiling. However, there are times when users might want to be profiled, such as to receive more relevant content during voice searches. In these cases, the ability to selectively edit or remove labels is important for reducing risks associated with profiling (e.g., due to mislabeling).
In our study, we found that only Google offered the option to fully opt out of profiling while also allowing users to manage their labels. It is important to note that Amazon's Profile Hub~\cite{amazon-profile-hub} does provide a method for users to declare their own labels. However, while these user-declared labels are added to the inferred ones, the inferred labels cannot be removed unless the user opts out entirely. The above tools align with existing literature that emphasizes the importance of giving users control over their data to enhance privacy and personalization~\cite{user-control-mitigation}.

\subsection{Privacy Policies vs. Our Results}
\label{sec:privacy_polocies}
\noindent \emph{Google Privacy Policy:}
An in-depth review of Google Assistant's privacy policy ~\cite{google-privacy-policy} indicates that personalized advertisements are fundamentally rooted in user-specific data, such as the data we provide to Google when creating accounts, our activity using Google services, and our location information. The policy clearly excludes sensitive details like race, religion, sexual orientation, and health from ad targeting.  It is important to highlight that all profiling activities we observed are allowed by the policy. Given the policy permits the use of all activities on Google services, it is possible to observe new demographic or interest profiling labels derived from user interactions with Google services in the future.

\emph{Amazon Privacy Policy.}
Upon reviewing Amazon's privacy policy~\cite{amazon-privacy-policy}, it is clear that personalized advertising is integral to their service, with ads being tailored based on a mixture of recommended products/services and personal user information. The policy elaborates on the extensive range of personal data collected, including basic identification and contact details, interaction logs, and personal identifiers. It is important to note our observed extent of profiling is less than the potential scope allowed in the policy.

\emph{Apple Privacy Policy.}
Apple's approach to user data for advertising purposes, as outlined in its privacy policy~\cite{apple-advertising}, reveals a commitment to leveraging user interactions for targeted advertising. Even if we did not observe any evidence of profiling, the policy acknowledges the profiling potential, which can happen in the future.

\subsection{Limitations}
\label{sec:limitations}
\noindent Our study faced multiple limitations that narrowed the depth and scope of our analysis, each presenting unique challenges:

    \emph{Scale of the experiments.} The requirement to create numerous new accounts for controlled experiments with voice assistants was the foremost challenge in conducting the experiments. Although some account creations did not need any phone numbers, some accounts needed phone numbers, and since we wanted to have completely independent accounts, we had to use a unique phone number for each account. This process was time-consuming, and significantly slowed down our research progress.
    
    \emph{Profile contamination.} Although we used different phone numbers for each account, certain identifiers such as IP addresses and device IDs remained unchanged across different experiments due to the reuse of the same devices for multiple experiments (and, to avoid IP-address-dependent profiling). This can be a source of contamination.
    
    \emph{Opaque box assumption and dynamism.} We are not testing static software, but cloud services that are constantly being updated, and therefore every experiment may be a test of a different system. This variability could provide an alternative explanation for observing different results or behaviors at different times.
    
    \emph{Voice characteristics.} We are assuming that voice characteristics are not used for profiling purposes, but previous research~\cite{Singh2019VoiceProfiling} has shown that there is potential for profiling also based on that. This suggests the possibility that our results could be influenced by the use of voice characteristics profiling. To mitigate this limitation, we used the same voice for all experiments.

    \emph{Undisclosed labels.} Our reliance on the visibility of labels for compliance with privacy legislation does not account for the potential of undisclosed profiling by voice assistants, which are essentially opaque boxes. Although we aim to provide a lower bound for profiling based on disclosed information, we acknowledge the limitation of not being able to detect 
    undisclosed profiling activities.
    
    \emph{Language specificity.} Another limitation of our study is the use of a specific language, English, for conducting experiments. The profiling behaviors and results may vary with other languages, which limits the generalizability of our findings for non-English speaking users.

Despite these limitations, our study provides valuable insights into the profiling behaviors of Google, Alexa, and Siri through voice interactions, laying the groundwork for future research to explore these areas further and advocate for greater transparency in voice assistant technologies.

\section{Conclusion}
\label{sec:conclusion}

In this study, we provided an overview of profiling practices of voice assistants. Through an extensive persona-based experimentation spanning \numexperiments{} experiments and \numqueries{} voice queries over a period of \nummonths{} months, we confirmed that Google Assistant and Amazon Alexa engage in user profiling based on voice interactions, while 
evidence of such profiling was not found for Apple Siri, despite Apple's disclosure in its privacy policy.

We observed distinct profiling methods between Google and Alexa, with Google employing a probabilistic approach that continuously refines demographic profiling labels based on user interactions, while Alexa adopts deterministic interest profiling labels primarily from explicit user commands within the Amazon shopping ecosystem. 
%
Our investigation also revealed different profiling behavior between voice and web interactions for Google, with lower profiling accuracy and longer profiling periods for the voice modality.
%
Last, while all services offer opt-out mechanisms, only Google allows users to selectively delete or amend labels.

Through this study we hope to motivate voice assistant companies to improve transparency and refine tools for mitigating misprofiling and empowering users to edit their profiles.

\bibliographystyle{ACM-Reference-Format}
\bibliography{main}

\appendix

\section{Query Generation Prompts}
\label{sec:query-generation-prompts}
All queries we used are also available. \footnote{https://anonymous.4open.science/r/PETS2025}

\subsection{Google Query Generation Prompt}
\label{sec:google-query-generation-prompt}

\subsubsection{Relationships Sample Query Generator Prompt}
Design 20 queries to ask a Google Voice Assistant that clearly identify me as a person that is single (Not married and not in a relationship). Ensure the questions reflect my specific relationship status accurately and distinctly. the queries should be simple and short, with the use of related key words.

\subsubsection{Household Income Sample Query Generator Prompt}
Design 20 queries to ask a Google Voice Assistant that clearly identify me as a person with a high income (not moderately high income and not average or lower income). Ensure the questions reflect my specific household income status accurately and distinctly. the queries should be simple and short, with the use of related key words.

\subsubsection{Education Sample Query Generator Prompt}
Design 20 queries to ask a Google Voice Assistant that clearly identify me as a person with a high school diploma. These queries should not suggest any level of education higher or lower than a high school diploma. Ensure the questions reflect my specific educational background accurately and distinctly. the queries should be simple and short, with the use of related key words.

\subsubsection{Homeownership Sample Query Generator Prompt}
Design 20 queries to ask a Google Voice Assistant that clearly identify me as a person that is renter (not homeowner). Ensure the questions reflect my specific homeownership status accurately and distinctly. the queries should be simple and short, with the use of related key words.

\subsubsection{Parenting Sample Query Generator Prompt}
Design 20 queries to ask a Google Voice Assistant that clearly identify me as a person that is parent of infants (not parent of toddlers, preschoolers, grade schoolers, or teenagers). Ensure the questions reflect my specific parenting status accurately and distinctly. the queries should be simple and short, with the use of related key words.

\subsection{Amazon Query Generation Prompt}
\label{sec:amazon-query-generation-prompt}

\subsubsection{Command Query Generator Prompt}
Imagine you are a person interested in [specific product category] products. Generate 30 simple and specific command queries for an Amazon Voice Assistant to show your interest. all queries must be either adding a product to cart or adding a product to shopping list. do not ask general queries. do not have any subscribe in your queries. we do not have any specific cart we have only “add to my cart” but we have add to my x list (it means for lists you can have names but for cart we have only add to cart command.

\subsubsection{General Query Generator Prompt}
Imagine you are a person interested in [specific product category] products. Generate 30 simple and specific queries for an Amazon Voice Assistant that express your interest in this category. Ensure the queries are general in nature, avoiding any command-like phrasing such as 'Add to cart' or 'Add to list.' Focus on inquiries that reflect curiosity, desire for information, or exploration of related products

\section{previous experiments in the previous paper version}
\label{sec:previous-experiments-in-the-previous-paper-version}
The details of previous experiments for the four research questions have been mentioned hereunder:
\subsection{Prepopulated Labels in Profiles}
\label{sec:Exploring_Prepopulated_Labels_in_Profiles_Appendix}

\begin{table}[t]
\begin{tabular}{llr}
\hline
\textbf{Category}                       & \textbf{Label}      & \multicolumn{1}{l}{\textbf{Count/30}} \\ \hline
\multirow{4}{*}{\textbf{Relationship}}  & Not Enough Info     & 16                                              \\ \cline{2-3} 
                                        & Single              & 8                                               \\ \cline{2-3} 
                                        & Married             & 4                                               \\ \cline{2-3} 
                                        & In a Relationship   & 2                                               \\ \hline
\multirow{3}{*}{\textbf{Education}}     & Not enough info     & 24                                              \\ \cline{2-3} 
                                        & Bachelor's degree   & 3                                               \\ \cline{2-3} 
                                        & High school diploma & 3                                               \\ \hline
\multirow{2}{*}{\textbf{Homeownership}} & Not enough info     & 7                                               \\ \cline{2-3} 
                                        & Homeowners          & 23                                              \\ \hline
\multirow{2}{*}{\textbf{Parenting}}     & Not enough info     & 4                                               \\ \cline{2-3} 
                                        & Not Parents         & 26                                              \\ \hline
\end{tabular}
\caption{Distribution of demographic tags assigned to new Google accounts during an initial profiling phase in Fall 2023.}
\label{tab:prepopulated_labels}
\end{table}

In Spring 2023, we found no pre-populated labels in 16 new Google accounts, with all labels showing ``Not enough info.'' However, in Fall 2023, many new Google accounts were created with pre-populated labels, notably ``Homeowners'' and ``Not Parents.'' The results are shown in Table \ref{tab:prepopulated_labels}.

During all our experiments with fresh accounts in Alexa and Siri, we have not seen preopopulated profiling labels.

\subsection{Characterization of Profiling Activities}
\subsubsection{Summary of Persona Experiments}

The study conducted a series of experiments to evaluate Google's profiling accuracy using both simple and complex personas. The personas were designed as follows:

\begin{itemize}
    \item \textbf{Persona 1:} ``Married'', ``Bachelor’s Degree''. Targeted the relationship and education demographic labels.
    \item \textbf{Persona 2:} ``Single'', ``High School Diploma''. Also targeted relationship and education labels.
    \item \textbf{Persona 3:} ``Married'', ``Homeowner'', ``Parent'', ``Bachelor’s Degree''. This complex persona targeted four demographic labels: relationship, homeownership, parenting, and education.
    \item \textbf{Persona 4:} ``Single'', ``Renter'', ``Not a Parent'', ``High School Diploma''. This persona similarly targeted the same four demographic labels.
\end{itemize}

\paragraph{Simple Persona Experiments:} 
In both Spring and Fall 2023, we tested the accuracy of Google’s profiling using Personas 1 and 2. The results, as shown in Table \ref{tab:google_simple_persona}, were consistent across both periods for the education label, but there were differences in profiling accuracy for the relationship label. The experiments in both periods show the evidence of profiling based on user interaction with Google assistants.

\paragraph{Complex Persona Experiments:}
In Fall 2023, we introduced complex personas (Personas 3 and 4) to explore how Google handles more detailed profiling. The results showed generally lower accuracy for relationship and education labels compared to simple personas. For the homeownership and parenting labels, accuracy varied significantly—one persona had nearly perfect accuracy, while the other showed none.

\paragraph{Accuracy Experiments with Repeated Training:}
To understand if the low accuracy in the complex persona experiments was due to insufficient training, we conducted additional experiments with repeated queries:

\begin{itemize}
    \item \textbf{Short-Term Repetition:} This involved repeating all training questions for each persona five times daily over one week. The results were mixed—Persona 4 showed an increase in relationship accuracy (from 30\% to 75\%), but Persona 3's relationship accuracy dropped to 0\%. Education accuracy for Persona 3 improved slightly (from 0\% to 25\%), while Persona 4's education accuracy decreased to 0\%.
    
    \item \textbf{Long-Term Repetition:} Conducted in January 2024, this experiment used the same Google accounts from Fall 2023. The accounts were re-exposed to the complex persona training questions after several months. This long-term repetition led to a significant improvement, with nearly all labels achieving 100\% accuracy.
\end{itemize}

The results, as shown in Table \ref{tab:google_complex_results}, indicate that while Google's profiling accuracy may initially be low for complex personas, repeated interactions, especially over a longer period, can substantially improve the accuracy of demographic labels.

\begin{table}[]
\begin{tabular}{cccc}
\hline
\textbf{Time}                         & \textbf{Persona} & \textbf{Relationship} & \textbf{Education} \\ \hline
\multirow{2}{*}{\textbf{Spring 2023}} & Persona 1        & 50\%                    & 62.5\%               \\ \cline{2-4} 
                                      & Persona 2        & 62.5\%                  & 100\%                \\ \hline
\multirow{2}{*}{\textbf{Fall 2023}}   & Persona 1        & 0\% (no tags)           & 66.67\%              \\ \cline{2-4} 
                                      & Persona 2        & 100\%                   & 100\%                \\ \hline
\end{tabular}
\caption{Comparison of accuracy metrics for Google's Simple Persona experiments in Spring and Fall 2023. The accounts were freshly created for each experiment. ``no tags'' indicates the absence of tags, with all profile labels concluding as ``Not Enough Info.''}
\label{tab:google_simple_persona}
\end{table}

\begin{table*}[]
\begin{tabular}{ccccccll}
\hline
\textbf{Complex Persona Experiments}                    & \textbf{Persona} & \textbf{Relationship}               & \textbf{Education}   & \textbf{Homeownership}  & \multicolumn{3}{c}{\textbf{Parenting}}      \\ \hline
\multirow{2}{*}{\textbf{Fall 2023 Initial New Accounts}}     & Persona 3    & 20\%                         & 0\% (no tags) & 90\%             & \multicolumn{3}{c}{0\% (inaccurate)} \\ \cline{2-8} 
                                                       & Persona 4    & 30\%                         & 30\%          & 0\% (inaccurate) & \multicolumn{3}{c}{100\%}            \\ \hline
\multirow{2}{*}{\textbf{Repeating 5x Daily New Accounts}}    & Persona 3    & 0\% (no tags and inaccurate) & 25\%          & 25\%             & \multicolumn{3}{c}{0\% (inaccurate)} \\ \cline{2-8} 
                                                       & Persona 4    & 75\%                         & 0\% (no tags) & 0\% (inaccurate) & \multicolumn{3}{c}{100\%}            \\ \hline
\multirow{2}{*}{\textbf{January 2024 Fall Initial Accounts}} & Persona 3    & 100\%                        & 100\%         & 100\%            & \multicolumn{3}{c}{0\% (inaccurate)} \\ \cline{2-8} 
                                                       & Persona 4    & 100\%                        & 100\%         & 100\%            & \multicolumn{3}{c}{100\%}            \\ \hline
\end{tabular}
\caption{Accuracy of label assignment for Google Complex Persona experiments across three scenarios (in percentage): initial assessment on fresh accounts (Fall 2023), repeated query exposure (5x daily for one week), and a follow-up evaluation (January 2024) on initially queried accounts. ``no tags'' indicates all labels were categorized as ``Not Enough Info'', while ``inaccurate'' denotes incorrect label assignment.}
\label{tab:google_complex_results}
\end{table*}

\begin{table}[t]
\begin{tabular}{clcc}
\hline
\multicolumn{1}{l}{\textbf{Query}} & \textbf{Persona} & \multicolumn{1}{l}{\textbf{Avg Accuracy}} & \multicolumn{1}{l}{\textbf{Mean Time (Days)}} \\ \hline
\multirow{2}{*}{\textbf{General}}             & Fashion      & 0                                          & N/A                                           \\ \cline{2-4} 
                                              & Video        & 0                                          & N/A                                           \\ \hline
\multirow{2}{*}{\textbf{Command}}             & Fashion      & 100\% ± 0\%                                    & 4.67 ± 0.58                                   \\ \cline{2-4} 
                                              & Video        & 100\% ± 0\%                                    & 5.33 ± 0.58                                   \\ \hline
\end{tabular}
\caption{Comparison of accuracy metrics for Alexa's general and command queries across Fashion and Video Personas in Summer 2023 and January 2024.}
\label{tab:alexa-general-command-old}
\end{table}

To analyze Alexa profiling accuracy we created two personas targeting the Video Entertainment and Fashion interest labels. We conducted our initial Amazon experiments during Summer 2024 utilizing a total of 12 new Amazon accounts, allocating six to each persona. The findings of these experiments are presented in Table ~\ref{tab:alexa-general-command-old}, showing similar results as depicted in section \ref{sec:impact_of_interaction_modality_on_user_profiling} showing that general questions do not results in profiling, while command-based queries result in 100\% profiling accuracy.

Experiments with simple personas linked to Google accounts also showed no indirect profiling by Google through Siri voice searches, suggesting Apple does not share user identifiers that lead to profiling on Google's end.

\subsection{Impact of Interaction Modality}
To evaluate the impact of different interaction modalities on user profiling, experiments were conducted across various modalities for Google and Alexa. These experiments compared the profiling accuracy and speed of voice versus web interactions using simple personas.

For Google, results summarized in Table~\ref{tab:modalities}, showed that web interactions were more effective for profiling, achieving 100\% accuracy for both two-label personas, while voice interactions had lower accuracy (56.25\% for Persona 1 and 81.25\% for Persona 2). Additionally, web interactions resulted in relatively faster tag changes (around 3.5 days) compared to voice interactions (around 4.9 days).

We studied the impact of voice and web interaction modalities on Amazon's user profiling, focusing on ``Fashion'' and ``Video'' Personas. For ``Fashion'', both modalities were tested, while for ``Video'', only voice commands were used due to the lack of a web equivalent. Results , summarized in Table~\ref{tab:modalities}, showed that web interactions achieved 100\% accuracy and relatively similar label changes compared to voice interactions. We were unable to investigate how different modalities affected Siri profiling due to not offering the search through a web interface.

\begin{table}[]
\begin{tabular}{cclcc}
\hline
\textbf{VA}         & \textbf{Mode}                       & \multicolumn{1}{c}{\textbf{Persona}} & \textbf{Accuracy} & \textbf{Time (Days)} \\ \hline
\multirow{4}{*}{\textbf{Google}} & \multirow{2}{*}{Voice} & Persona 1                        & 56.25\% ± 8.84\%           & 4.915 ± 0.83              \\ \cline{3-5} 
                                 &                                              & Persona 2                        & 81.25\% ± 17.68\%          & 4.895 ± 0.39              \\ \cline{2-5} 
                                 & \multirow{2}{*}{Web}   & Persona 1                        & 100\% ± 0\%                & 3.5 ± 0.71                \\ \cline{3-5} 
                                 &                                              & Persona 2                        & 100\% ± 0\%                & 3.435 ± 0.27              \\ \hline
\multirow{2}{*}{\textbf{Alexa}}  & Voice                  & Fashion                                   & 100 ± 0                & 4.67 ± 0.58               \\ \cline{2-5} 
                                 & Web                    & Fashion                                   & 100\% ± 0\%                & 3.5 ± 0.71                \\ \hline
\end{tabular}
\caption{Comparison of mean profiling accuracy and the mean time required for accurate label convergence across various modality types for Google and Alexa voice assistants.}
\label{tab:modalities}
\end{table}

\end{document}